\documentclass[
twocolumn,
showkeys,
superscriptaddress,
amsmath,amssymb,
aps,
nofootinbib,
floatfix
]
{revtex4}

\usepackage{graphicx}
\usepackage{dcolumn}
\usepackage{bm}
\usepackage{float}
\usepackage[colorlinks=true,urlcolor=blue,breaklinks=true]{hyperref}
\usepackage{upgreek}
\usepackage{bibunits}

\def\bibsection{\section{References}} 

\begin{document}

\begin{bibunit}
\title{Transmission phase read-out of a large quantum dot in a nanowire interferometer}

\author{Francesco Borsoi}
\affiliation{QuTech and Kavli Institute of Nanoscience, Delft University of Technology, 2600 GA Delft, The Netherlands}

\author{Kun Zuo}
\affiliation{RIKEN Center for Emergent Matter Science (CEMS), Wako 351-0198, Saitama, Japan}

\author{Sasa Gazibegovic}
\author{Roy L. M. Op het Veld}
\author{Erik P. A. M. Bakkers}
\affiliation{Department of Applied Physics, Eindhoven University of Technology, 5600 MB Eindhoven, The Netherlands}

\author{Leo P. Kouwenhoven}
\affiliation{QuTech and Kavli Institute of Nanoscience, Delft University of Technology, 2600 GA Delft, The Netherlands}
\affiliation{Microsoft Quantum Lab Delft, 2600 GA Delft, The Netherlands}

\author{Sebastian Heedt}
\email{Sebastian.Heedt@Microsoft.com}
\affiliation{QuTech and Kavli Institute of Nanoscience, Delft University of Technology, 2600 GA Delft, The Netherlands}
\affiliation{Microsoft Quantum Lab Delft, 2600 GA Delft, The Netherlands}
\date{\today}

\begin{abstract}
Detecting the transmission phase of a quantum dot via interferometry can reveal the symmetry of the orbitals and details of electron transport. Crucially, interferometry will enable the read-out of topological qubits based on one-dimensional nanowires. However, measuring the transmission phase of a quantum dot in a nanowire has not yet been established. Here, we exploit recent breakthroughs in the growth of one-dimensional networks and demonstrate interferometric read-out in a nanowire-based architecture. In our two-path interferometer, we define a quantum dot in one branch and use the other path as a reference arm. We observe Fano resonances stemming from the interference between electrons that travel through the reference arm and undergo resonant tunnelling in the quantum dot. Between consecutive Fano peaks, the transmission phase exhibits phase lapses that are affected by the presence of multiple trajectories in the interferometer. These results provide critical insights for the design of future topological qubits.
\end{abstract}
\maketitle

\section{Introduction}
\begin{figure}[hbt!]
\includegraphics{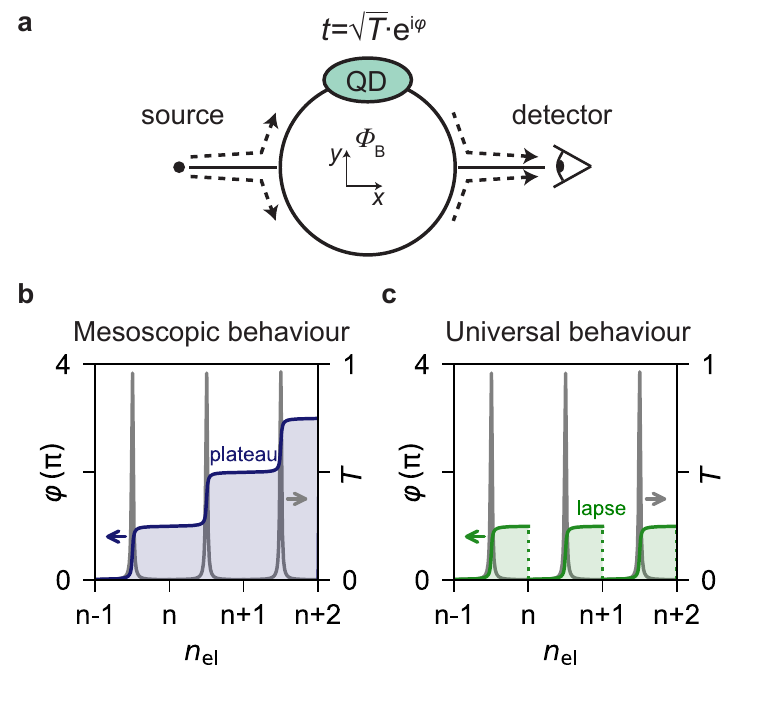}
\caption{Mesoscopic and universal phase behaviours. \textbf{a}~The minimum setup to study the transmission phase via a quantum dot (in light green) is a two-path interferometer. \textbf{b}, \textbf{c}~Transmission phase $\varphi$ and probability $T$ as a function of the electron number $n_{\mathrm{el}}$ in a quantum dot. A Breit-Wigner function describes each of the resonances (in grey). \textbf{b}~The mesoscopic regime: phase plateaus in the Coulomb valleys appear at $0$ and $\uppi$. \textbf{c}~The universal regime: phase lapses occur between transmission resonances.}
\label{fig:Figure 1}
\end{figure}

Similar to a light wave, an electron wave acquires a phase when interacting with a scattering centre. Studying this effect requires an interferometer with phase-coherent transport such as semiconducting or metallic rings~\cite{WEB1985, BAC1999, RUS2008}. In these nanostructures, the phase difference between the two paths ($\Delta\varphi$) can be tuned by a magnetic flux via the Aharonov-Bohm (AB) effect:
\begin{equation}
\Delta\varphi = 2 \uppi \frac{\mathit{\Phi_{\mathrm{B}}}}{\mathit{\Phi_{\mathrm{0}}}}
\label{eq: AB relation}
\end{equation}
with $\mathit{\Phi}_\mathrm{B}$ the magnetic flux through the interferometer and $\mathit{\Phi}_\mathrm{0}=h/e$ the flux quantum.\\
When the scattering centre is a quantum dot (QD), as depicted in Fig.~\ref{fig:Figure 1}a, the transmission phase $\varphi$ provides information complementary to the transmission probability $T = |t|^2$, with $t$ the transmission amplitude $ t = \sqrt{T} \mathrm{e}^{\mathrm{i}\varphi}$. It can reveal insights into microscopic details of electron transport and into the spatial symmetries of the orbitals~\cite{LEE1998, ORE2007, SIL200, VAN2006}.
~\\
Recently, theoretical proposals suggested using interferometry as a read-out method of topological qubits, where quantum information is encoded in the electron parity of Majorana modes in semiconducting-superconducting nanowires~\cite{KIT2001, LUT2010, ORE2010, ALI2012, VIJ2016, PLU2017, KAR2017}. Here, opposite qubit states are characterised by different transmission phases similar to the mesoscopic phase behaviour observed in few-electron quantum dots (Fig.~\ref{fig:Figure 1}b)~\cite{FU2010, DRU2018, HEL2018, AVI2005}. 
When Majorana modes are absent, the phase is expected to exhibit the universal behaviour detected in many-electron quantum dots. In these systems, abrupt phase lapses break the simple parity-to-phase relation (Fig.~\ref{fig:Figure 1}c)~\cite{YAC1995, SCH1997, ORE2007, KAR2007, AVI2005, EDL2017}.\\
Despite the critical application in topological qubits, the phase read-out of a quantum dot in a nanowire interferometer has not been demonstrated yet. 
While pioneering works employed two-dimensional electron gases~\cite{YAC1995, SCH1997, AVI2005, EDL2017}, here we take advantage of the recent advances in the growth of nanowire networks~\cite{CAR2014, GAZ2017} and demonstrate interferometric read-out of a quantum dot defined in a nanowire. Our findings provide crucial insights for future topological qubits based on hybrid one-dimensional nanowire systems.

\section{Results}
\subsection{Cotunnelling Aharonov-Bohm interference}
\begin{figure}[htb]
\includegraphics{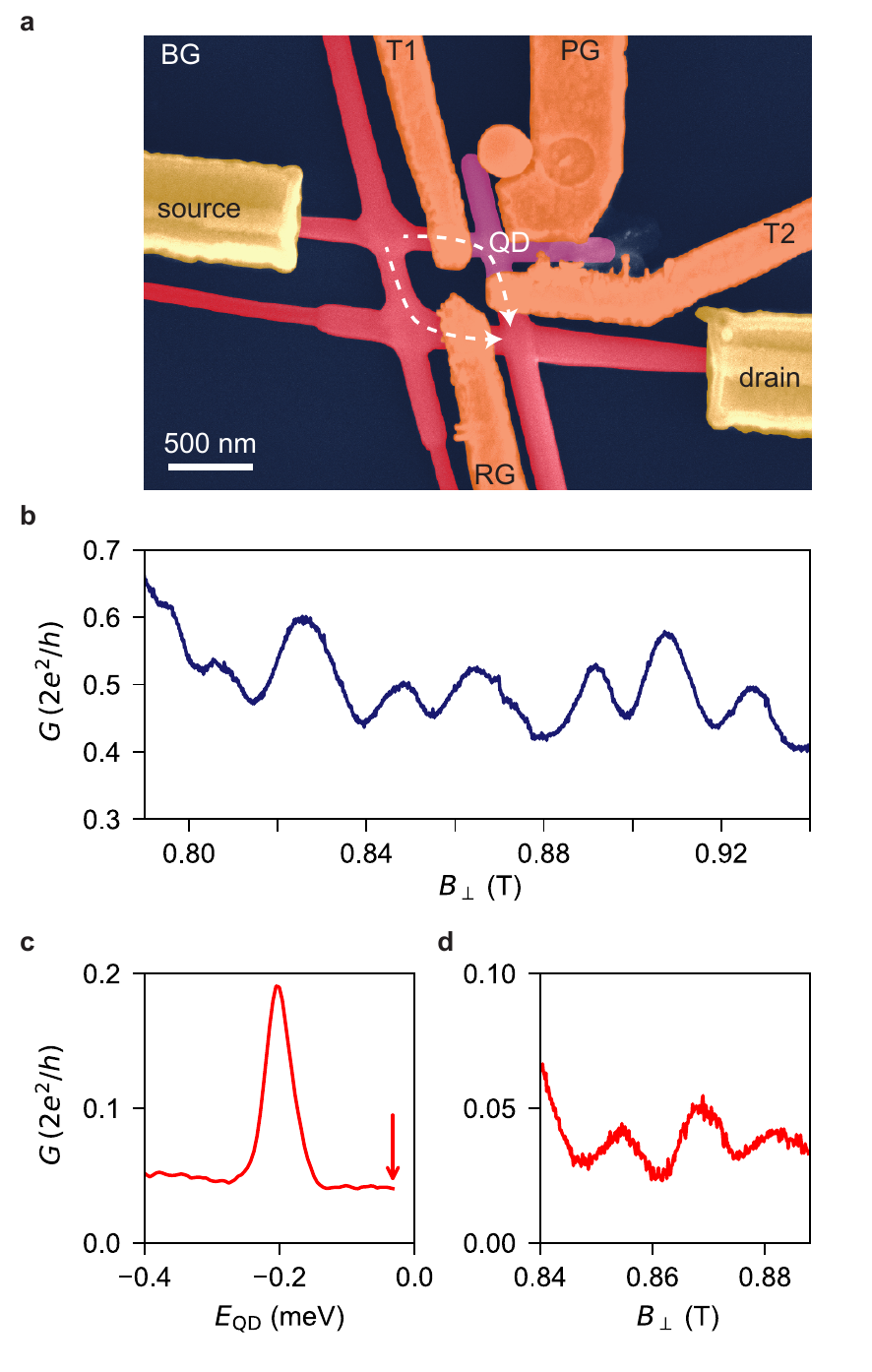}
\caption{Aharonov-Bohm oscillations in an InSb nanowire network. \textbf{a}~False-colour scanning electron micrograph of the device: in red the nanowire network, in gold the leads, in orange the gates and in pink the quantum dot region. An additional illustration and a schematic of the device are shown in the Methods. \textbf{b}~Conductance at zero bias voltage $G (V_{\mathrm{SD}} = 0)$ as a function of the perpendicular field $B_{\perp}$ in the open regime (i.e.,\ with no QD defined) manifesting AB oscillations. \textbf{c}~$G (V_{\mathrm{SD}} = 0)$ vs.\ $E_{\mathrm{QD}}$ (the dot electrochemical potential) when the quantum dot is defined. \textbf{d}~$G (V_{\mathrm{SD}} = 0)$ vs.\ $B_{\perp}$ when the dot is in the cotunnelling regime (cf.\ Coulomb valley indicated by the red arrow in panel \textbf{c}).}
\label{fig:Figure 2}
\end{figure}

Our device is shown in Fig.~\ref{fig:Figure 2}a and consists of a hashtag-shaped network of hexagonal InSb nanowires of high crystalline quality~\cite{GAZ2017}. In the top-right arm, negative voltages ($V_{\mathrm{T1}}$ and $V_{\mathrm{T2}}$) on the top gates, T1 and T2, create two tunnel barriers that define an X-shaped quantum dot (pink region). The voltage on the plunger gate PG ($V_{\mathrm{PG}}$) tunes its electron occupation. Likewise, the transmission in the bottom-left branch -- the reference arm -- can be varied from pinch-off to the open regime by adjusting the voltage $V_{\mathrm{RG}}$ on the reference gate (RG). The $p$-doped-Si/SiO$_x$ substrate allows global back-gate (BG) functionality. A DC bias voltage with a small AC excitation, $V_{\mathrm{SD}} + \delta V_{\mathrm{AC}}$, is applied between source and drain, yielding a current $I + \delta I_{\mathrm{AC}}$. Both the DC current and the differential conductance $G = \delta I_{\mathrm{AC}} / \delta V_{\mathrm{AC}}$ are measured in a dilution refrigerator with an electron temperature of $T_{\mathrm{el}} \sim 35\,$mK at its base temperature.
~\\
When the QD is not defined, the conductance at zero bias voltage displays Aharonov-Bohm oscillations as a function of the magnetic field perpendicular to the substrate ($B_{\perp}$) with period $\Delta B_{\perp} \sim 16 - 20\,$mT (Fig.~\ref{fig:Figure 2}b). This periodicity corresponds to a loop area of $\mathit{\Phi}_\mathrm{0}/\Delta B_{\perp} \sim 0.21 - 0.26\, \mathrm{\upmu m^2}$, which is consistent with the actual area of the device of $\sim 0.23 \,\mathrm{\upmu m^2}$ measured up to the centre of the nanowires.\\
When the quantum dot is defined, we adjust the plunger-gate voltage between two resonances, as indicated by the red arrow in Fig.~\ref{fig:Figure 2}c, where the horizontal axis is the QD electrochemical potential ($E_{\mathrm{QD}} = e \cdot \alpha \cdot V_{\mathrm{PG}}$, with $\alpha$ the lever arm and $e$ the electron charge). In this regime, the electron-electron repulsion in the dot suppresses the current almost completely, which is known as Coulomb blockade. Transport is then allowed only via virtual, higher-order processes. At zero bias, elastic cotunnelling is predominant and its phase coherence is critical for parity-protected read-out schemes of Majorana wires~\cite{PLU2017, KAR2017}.\\
When we balance the current distribution in the two arms of our device, the AB oscillations in the cotunnelling regime become visible with an amplitude of $\sim 20 - 30 \% $ of the average conductance (Fig.~\ref{fig:Figure 2}d). The large visibility demonstrates that cotunnelling across the large Coulomb-blockaded dot is phase-coherent, fulfilling a fundamental requirement of future parity read-out circuits. 

\subsection{From Coulomb to Fano resonances}
\begin{figure}[hbt!]
\includegraphics{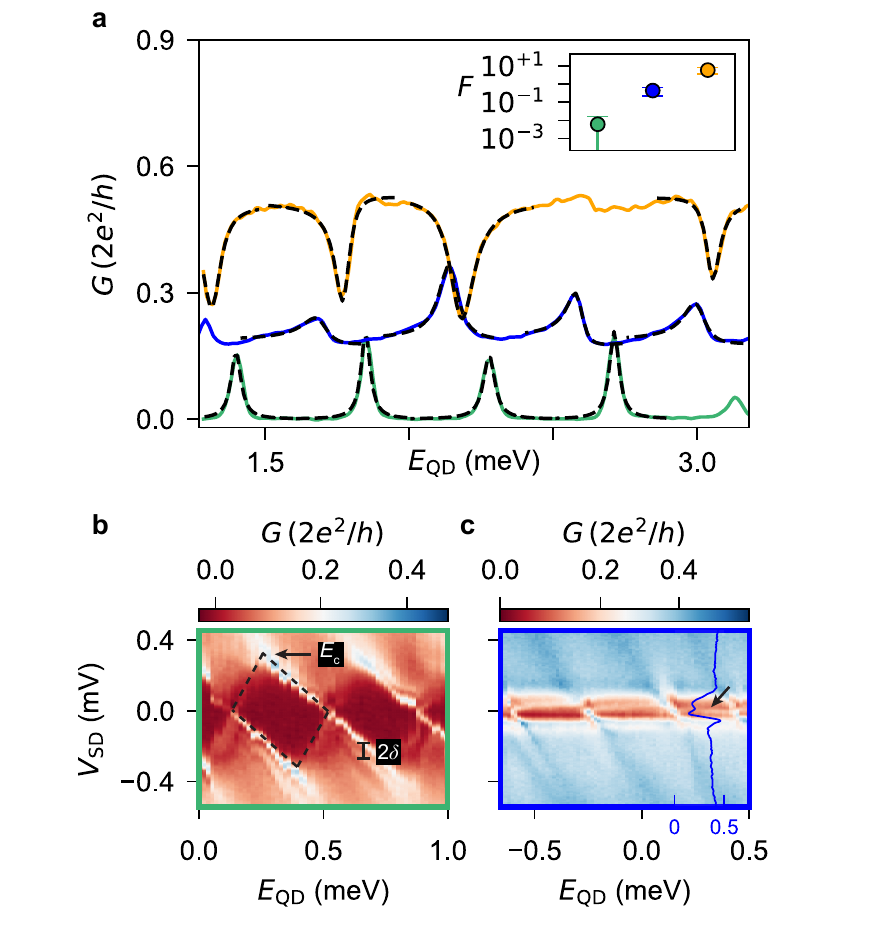}
\caption{From Coulomb to Fano resonances. \textbf{a}~Differential conductance $G$ as a function of $E_{\mathrm{QD}}$ with the reference arm fully pinched-off (green trace), partially conducting (blue trace) and transparent (orange trace). Dashed lines are best fits. Inset: Fano parameter $F = t_{\mathrm{ref}}/\sqrt{J_\mathrm{L} J_\mathrm{R}}$, averaged across four peaks in each of the three regimes. \textbf{b}, \textbf{c}~$G$ versus $E_{\mathrm{QD}}$ and $V_{\mathrm{SD}}$ in the first and second regime, respectively. The blue line-cut in \textbf{c} is taken at $E_{\mathrm{QD}} = 0.32\,$meV, the blue values on the horizontal axis refer to conductance $G$ in $2e^2/h$. In \textbf{b}, $E_{\mathrm{c}}$ indicates the charging energy (at the apex of the diamond) and $\delta$ denotes the level spacing due to quantum confinement.}
\label{fig:Figure 3}
\end{figure}

In order to characterize the quantum dot, we first pinch off the reference arm. The green trace in Fig.~\ref{fig:Figure 3}a displays a series of nearly equally spaced conductance peaks stemming from tunnelling via the dot. Their separation is also known as the addition energy and arises from two effects: the quantum confinement and the Coulomb interaction~\cite{KOU2001}. In a large dot, the second effect dominates over the first, leading to a series of peaks that are equidistant~\cite{KOU2001, IHN2010}. From the bias spectroscopy in Fig.~\ref{fig:Figure 3}b, we estimate the Coulomb charging energy $E_{\mathrm{c}} = e^2/C \sim 0.35 - 0.45 \,$meV (with $C$ the overall capacitance) and the level spacing due to confinement $ \delta \sim 0.020-0.035\,$meV. We evaluate the first parameter from the size of the diamonds in bias voltage, and the second from the separation between the lines that confine the Coulomb diamonds and the lines that are due to the excited states. The large ratio of $E_{\mathrm{c}} / \delta \gg 1$ indeed arises from the large size of the dot, which is designed to be comparable with the typical micron-long semiconducting-superconducting dots of near-future explorations~\cite{ALB2016, VAN2018, SHE2018}. Assuming a typical open-channel electron density of $2 \cdot 10^{17}\,\mathrm{cm^{-3}}$~\cite{PLI2013} and the dot volume of $ 1.4 \cdot 10^{-2} \, \mathrm{\upmu m ^3}$, we estimate the maximum number of electrons on the QD to be $\sim 1-3 \cdot 10^3$.\\
We now start to activate transport in the reference arm.
Upon increasing its transparency, the Coulomb peaks first evolve into the asymmetric peaks of the blue trace and then into the dips in the orange one of Fig.~\ref{fig:Figure 3}a. 
The variation of their line-shapes stems from the Fano effect, a phenomenon observed in multiple contexts in physics: from Raman scattering~\cite{CER1973, GUP2003} to photon absorption in quantum-well structures~\cite{FAI1997, SCHM1997}, from transport in single-electron transistors~\cite{GOR2000} to Aharonov-Bohm interferometers~\cite{KOB2002, AHA2006, HUA2015, RYU1998, KOB2003, KAT2004}.\\
The effect originates from the interference between two partial waves: one is undergoing a resonant scattering and the other is travelling through a continuum of states. In our experiment, the first is mediated by the discrete dot spectrum provided by Coulomb blockade and confinement, and the second by the continuum of the density of states in the reference path. Bias spectroscopy with the reference path being partially conducting -- similarly to the blue trace in Fig.~\ref{fig:Figure 3}a -- shows Fano peaks extending into the Coulomb valleys at $V_{\mathrm{SD}} \sim 0\,$mV (cf.\ black arrows in Fig.~\ref{fig:Figure 3}c). To the best of our knowledge, this is the first observation of Fano physics in a nanowire-based interferometer.\\
To distinguish the three regimes of Fig.~\ref{fig:Figure 3}a, we fit the line-shapes of the peaks using a generalized Fano model~\cite{AHA2006}. The relevant ingredients are the coupling terms between the dot and the two leads ($J_{\mathrm{L}}$ and $J_{\mathrm{R}}$), the transmission through the reference arm ($t_{\mathrm{ref}}$) and the magnetic flux through the ring ($\mathit{\Phi}_{\mathrm{B}}$). A schematic illustration and more information are shown in the Methods and the result of the fits are listed in Supplementary Tables~1-3.\\
We extract the Fano parameter $F=t_{\mathrm{ref}}/\sqrt{J_{\mathrm{L}} J_{\mathrm{R}}}$ from each peak (or dip). The inset of Fig.~\ref{fig:Figure 3}a shows that the averages of $F$ across each trace extend over three orders of magnitude, reflecting the large tunability of the device.
\begin{figure}[hbt!]
\includegraphics{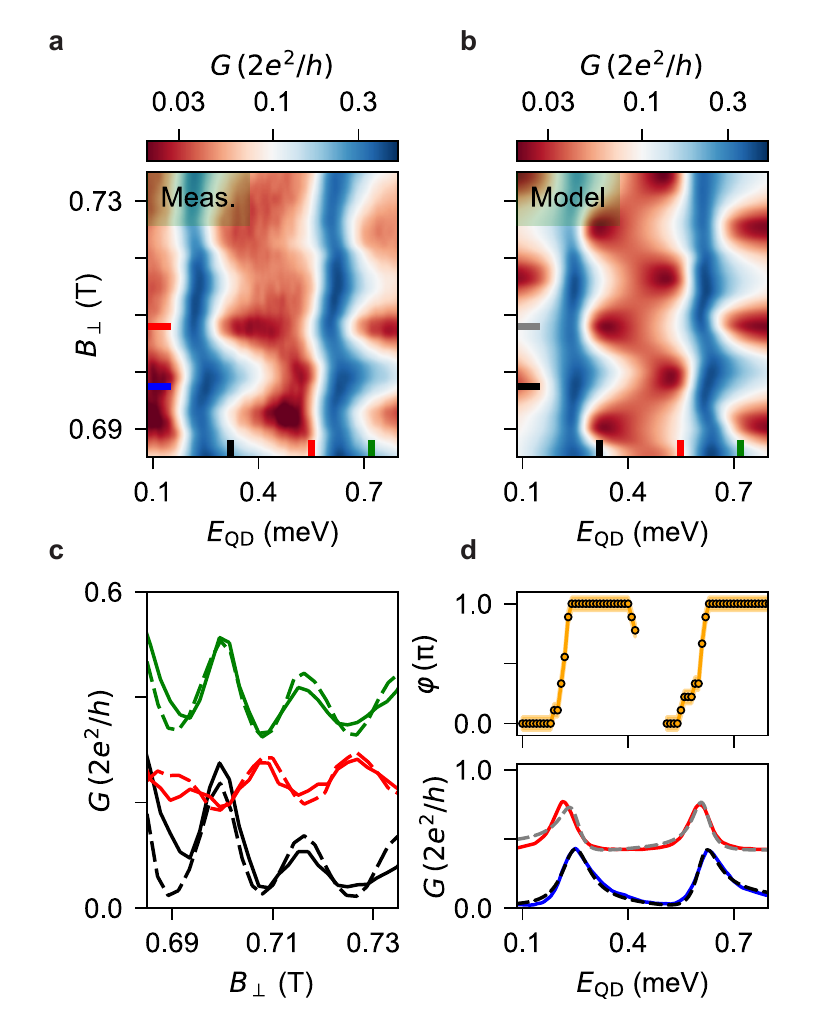}
\caption{The universal phase behaviour. \textbf{a}~$G$ versus $E_{\mathrm{QD}}$ and $B_{\perp}$ at zero bias voltage, describing the evolution of two adjacent Coulomb peaks as a function of magnetic field. Data are taken at back-gate voltage $V_\mathrm{BG} = 1.5\,$V. \textbf{b}~Fit of the data in panel \textbf{a}. \textbf{c}~Solid and dashed traces are vertical line-cuts from panels \textbf{a} and \textbf{b}, respectively, indicated by black, red, and green lines. The three pairs of curves are displaced by an offset of $0.15 \cdot 2e^2/h$ for clarity. \textbf{d}~Bottom panel: solid and dashed lines are horizontal line-cuts of panels \textbf{a} and \textbf{b}, respectively, at the positions indicated by the blue/black and red/grey lines. Top panel: transmission phase $\varphi$ extracted from the AB pattern. The shaded region indicates the error bars that stem from the uncertainty in extracting the oscillation maxima that is $\sim 1-2\,$mT.}
\label{fig:Figure 4}
\end{figure}
\subsection{The universal phase behaviour}
Upon sweeping the magnetic field, the Fano line-shapes vary periodically owing to the Aharonov-Bohm effect. In particular, Fig.~\ref{fig:Figure 4}a shows that two adjacent Fano resonances evolve in-phase.\\
We use the model described above to fit both peaks as a function of magnetic field, and we illustrate the result in Fig.~\ref{fig:Figure 4}b. The model captures well the main features of the experimental data, and the good agreement is visible in the line-cuts presented in Fig.~\ref{fig:Figure 4}c. Here, the three traces are taken at the positions denoted by the black, red, and green lines in both panels a and b. 
A $\uppi$-shift in the AB oscillations is visible between both the black and red as well as the red and green traces. The complete evolution of the phase $\varphi$ as a function of $E_{\mathrm{QD}}$ is extracted by tracking the maximum of the AB pattern and shown in the top panel of Fig.~\ref{fig:Figure 4}d. In the bottom panel, we present horizontal line-cuts of Figs.~\ref{fig:Figure 4}a and \ref{fig:Figure 4}b at the positions indicated by the coloured lines.\\
Here, we observe two main features: a phase variation of $\uppi$ at the resonances over an energy scale similar to the broadening of the peaks, and a phase lapse in the Coulomb valley. These are distinctive features of the universal phase behaviour and are consistent with the in-phase evolution of the two adjacent CPs in Fig.~\ref{fig:Figure 4}a~\cite{YAC1995, SCH1997, ORE2007, KAR2007, AVI2005}. The observation of the universal rather than the mesoscopic behaviour can be explained by taking a look at the energy scales of the transport. In our measurement, the typical dot coupling energy ($\mathit{\Gamma} = \sqrt{|J_\mathrm{L} J_\mathrm{R}|} \sim 0.1\,$meV) is a few times larger than the level spacing in the dot ($\delta \sim 0.02-0.035\,$meV). Therefore, tunnelling occurs via multiple dot-levels, a condition for which theory predicts the observation of the universal behaviour~\cite{KAR2007, ORE2007}.\\
Because previous experiments focused on the single-level regime ($\mathit{\Gamma} < \delta$)~\cite{SCH1997, YAC1995, AVI2005} and in the crossover ($\mathit{\Gamma} \sim \delta$)~\cite{EDL2017}, finding both phase lapses and phase plateaus, our investigation in the fully multi-level regime ($\mathit{\Gamma}\gg \delta$) seems to complete the complex dot-interferometry puzzle.
\subsection{Multi-path transport effects}
\begin{figure}[!]
\includegraphics{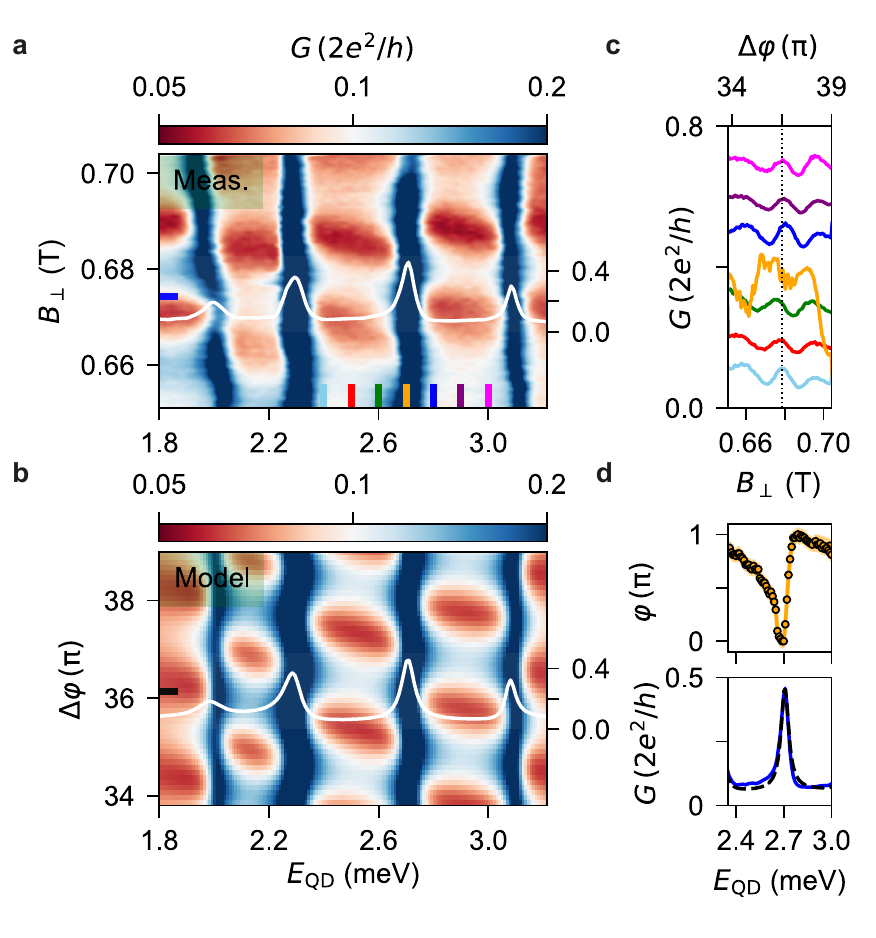}
\caption{Multi-path transport effects. \textbf{a}~$G$ vs.\ $E_{\mathrm{QD}}$ and $B_{\perp}$ exhibiting the evolution of four CPs. Data are taken at back-gate voltage $V_\mathrm{BG} = -1.5\,$V. \textbf{b}~Calculated conductance assuming a multi-path interferometer, details are reported in the Methods. The white traces in \textbf{a} and \textbf{b} correspond to the values indicated by the horizontal lines, and the vertical axes refer to conductance $G$ in $2e^2/h$. \textbf{c}~Vertical line-cuts of panel \textbf{a} showing the evolution of AB oscillations across a charge transition in the QD. Traces are displaced by $0.1\cdot 2e^2/h$ for clarity, except for the orange one taken on resonance. \textbf{d}~Top panel: the trend of the AB maxima across the third CP. The shaded region indicates the error bars stemming from the uncertainty in extracting the oscillation maxima that is $\sim 1-2\,$mT. Bottom panel: solid and dashed lines are horizontal line-cuts of panels \textbf{a} and \textbf{b}, respectively, at the position indicated by the blue/black lines.}
\label{fig:Figure 5}
\end{figure}

In the following, we highlight that an optimal read-out of the transmission phase requires an interferometer close to the one-dimensional limit in the sense that it shoud comprise thin nanowires enclosing a relatively large hole.\\
For a gate configuration different from the previous regime, the transmission phase varies smoothly between several pairs of adjacent CPs. Here, the phase displays a behaviour in between the universal and the mesoscopic regimes (Fig.~\ref{fig:Figure 5}a). 
The two configurations differ in the back-gate voltage that has been lowered from $V_{\mathrm{BG}} = 1.5\,$V in Fig.~\ref{fig:Figure 4} to $V_{\mathrm{BG}} = -1.5\,$V in Fig.~\ref{fig:Figure 5}. 
Voltages on the tunnel gates are also re-adjusted to retain a similar transmission, whereas the plunger gate remains at $V_{\mathrm{PG}}\sim 0\,\mathrm{V}$. We estimate a reduction of the electron density by no more than $\sim 20\% $ compared to the first case, leaving the dot still in the many-electron regime (see Supplementary Figure~2).\\
In Fig.~\ref{fig:Figure 5}a, we show a color map of $G$ vs.\ $E_{\mathrm{QD}}$ and $B_{\perp}$, exhibiting the evolution of 4 CPs. The red features in the cotunnelling regions oscillate as a function of $B_{\perp}$ owing to the AB effect. Several vertical line-cuts are shown in Fig.~\ref{fig:Figure 5}c. The maxima of the AB oscillations around $B_{\perp} \sim 0.68\,$T are converted into transmission phase via the magnetic field period (here $\Delta B_{\perp} = 19\,$mT) and displayed in the top panel of Fig.~\ref{fig:Figure 5}d. In the bottom panel of Fig.~\ref{fig:Figure 5}d, we show a horizontal line-cut taken at the position indicated by the blue line in Fig.~\ref{fig:Figure 5}a. Similar to the data in Fig.~\ref{fig:Figure 4}, the phase exhibits a $\sim \uppi$ variation concomitant with the peak in the conductance.
However, the phase lapse in the Coulomb valley is replaced by a smooth evolution. This slow phase variation is not universal, but depends on the specific gate setting.\\
We interpret this anomaly as a consequence of the relatively large width of the nanowires ($\sim 100 - 150\,$nm). Microscopically, we speculate that consecutive charge states might not couple to the same loop trajectory. The presence of at least two paths gives rise to beatings in the magneto-conductance that conceal the evolution of the transmission phase~\cite{AHA2006}. Our interpretation is well-supported by the large width of the Aharonov-Bohm peak in the Fourier spectrum shown in Supplementary Figure~3.\\
While in reality multiple trajectories could couple to each QD orbital, we reproduce our observation in the model by linking each resonance to a possibly different AB periodicity. This simple assumption enables to capture the main features of the measurement (Fig.~\ref{fig:Figure 5}b).\\
The coexistence of the two distinct phase behaviours (Fig.~\ref{fig:Figure 4} vs.\ Fig.~\ref{fig:Figure 5}) in the same mesoscopic device is hard to fully explain, and might be correlated with the exact coupling mechanism between the dot orbitals and the leads.

\section{Discussion}
In summary, we report interferometric measurements on a quantum dot embedded in a network of four conjoint InSb nanowires. The observation of pronounced quantum interference in the cotunnelling regime and the presence of Fano resonances suggest that interferometry is a viable tool for parity read-out of future topological qubits in nanowire networks. Theory suggests that the transmission probability of a semiconducting-superconducting quantum dot in the topological regime should exhibit phase plateaus~\cite{FU2010, DRU2018}. However, transmitting channels other than the teleportation via Majorana bound states were not taken into account. In experiments, extended topologically trivial modes without an underlying topological bulk phase can mimic Majoranas. Hence, quasiparticle transport via these modes might offer parallel paths to the Majorana teleportation~\cite{WHI2019, PRA2019}. Altogether, these can cause phase lapses that hinder the simple correspondence between the transmission phase and the electron parity. We conclude by remarking that future interferometers for parity-state discrimination via phase read-out should be designed with a large ratio between circumference and nanowire diameter.

\section{Methods}
\noindent \textbf{Device fabrication:}\\
InSb networks are grown by combining the bottom-up synthesis of four monocrystalline nanowires and the accurate positioning of the nanowire seeds along trenches on an InP substrate. Further details on the nanowire growth are presented in refs.~\onlinecite{CAR2014} and~\onlinecite{GAZ2017}.\\
After the growth, we transfer nanowire networks from the InP growth chip onto a $p$-doped Si/SiO$_x$ substrate (oxide thickness of $285\,$nm) using a mechanical nanomanipulator installed in a scanning electron microscope. Ti/Au contact leads are patterned using electron-beam lithography and e-gun evaporation, following surface treatment of the InSb for $30$ minutes in a sulfur-rich ammonium polysulfide solution diluted in water ($1:200$) at $60^{\circ}$C. The devices are covered with $\sim 30\,$nm of sputtered Si$_x$N$_y$ acting as a gate dielectric. The second layer of Ti/Au electrodes is patterned and evaporated to define the top gates. The chip is then diced, mounted and bonded onto a commercial printed circuit board.\\~\\

\noindent \textbf{Transport measurements:}\\
The device is cooled down in a dry dilution refrigerator equipped with a $6$-$2$-$2\,$T vector magnet. The base electron temperature is $T_{\mathrm{el}} \sim 35\,$mK. Conductance across the device is measured via a standard low-frequency lock-in technique at an AC signal amplitude of $\delta V_{\mathrm{AC}} \sim 20\,\upmu$V. The data presented in the main text and in Supplementary Figures~2, 3, and 4 are taken from a single device. In Supplementary Figures~5 and 6, we present data taken from a second and third device, respectively. The AC conductance in Figs.~\ref{fig:Figure 2}, \ref{fig:Figure 3}, \ref{fig:Figure 4}, and Supplementary Figure~4 was corrected for a constant offset that was later identified to arise from the setup.\\~\\

\begin{figure*}[hbt!]
\includegraphics{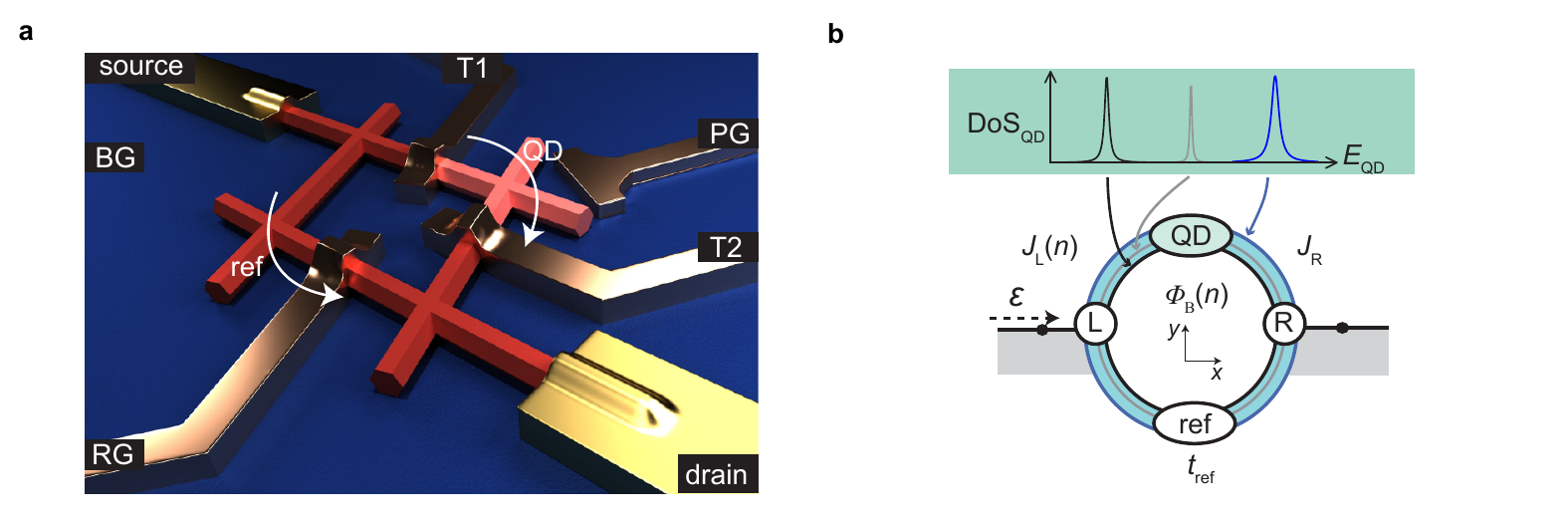}
\caption{The multi-path interferometer. \textbf{a}~Illustration of the device: in red the nanowire, in gold the leads, in copper the gates and in pink the quantum dot. \textbf{b}~Schematic of the device: the quantum dot exhibits a density of states (DoS$_{\textrm{QD}}$) comprised of discrete levels with distinct energy broadening. In the model, we assume that the dot states might couple to different interferometer trajectories that are sketched as rings of different colour.}
\label{fig:Figure 6}
\end{figure*}

\noindent \textbf{Model of the Aharonov-Bohm interferometer:}\\
The Landauer formula ($G = (2e^2/h) \cdot T$) connects the single-channel conductance of the system with the transmission probability $T$. In Fig.~\ref{fig:Figure 6}, we show a schematic of the multi-path Aharonov-Bohm interferometer (a simple generalization of the single-path counterpart) next to an illustration of the actual device. The QD electrochemical potential ladder is represented as a series of discrete states, separated by the charging energy $E_{\mathrm{c}} = e^2/C$~\cite{IHN2010}. For the single-path case, hopping terms $J_{\mathrm{L}}$ and $J_{\mathrm{R}}$ couple the source (L) and drain (R) to the QD, respectively, with the Aharonov-Bohm phase included in $J_\mathrm{L} = j_\mathrm{L} \cdot \exp(\mathrm{i} 2 \uppi \mathit{\Phi}_{\mathrm{B}}/\mathit{\Phi}_{\mathrm{0}})$, and $j_{\mathrm{L}}$ and $J_{\mathrm{R}}$ being real parameters. $\mathit{\Phi}_{\mathrm{B}}$ is the magnetic flux through the loop. When multiple-path are considered, we define the phase of $J_\mathrm{L}$ as $2 \uppi \mathit{\Phi}_{\mathrm{B}}/ \mathit{\Phi}_{\mathrm{0}} [1 + x(n)]$, with the parameter $x(n)$ distinct for every CP.\\
The reference site has a slowly varying spectrum that we will assume for simplicity to be constant. The leads are assumed to be one-dimensional (lattice constant $a$) with hopping matrix elements $-J$ and a typical energy dispersion of $\varepsilon = -2J\cos(ka)$. The resulting transmission probability $T$ through the AB interferometer is~\cite{AHA2006}:
\begin{equation} 
T = \frac{4 |S_{\mathrm{LR}}|^2 \cdot \sin^2(ka)}{\left||S_{\mathrm{{LR}}}|^2 - (S_{\mathrm{LL}} + e^{-ika})(S_{\mathrm{RR}} + e^{-ika})\right|^2}
\label{eq:1}
\end{equation}
with
\begin{equation} 
S_{XY} = \sum_{n = 0}^{N} {\frac{J_{\mathrm{X}} (n) J_{\mathrm{Y}} (n)^*}{J\left(\varepsilon - E_{\mathrm{QD}} + E_\mathrm{n} \right)}} + t_{\mathrm{ref}} 
\label{eq:2}
\end{equation}
where $E_\mathrm{n}$ represent the positions of the $N$ levels relevant in the tunnelling process on the $E_{\mathrm{QD}}$-axis. For the fit of our results, we add an offset to eq.~\eqref{eq:1} to capture the incoherent contribution of the current through the device.

\section{Data availability}
The data that support the findings of this study are available at \url{https://doi.org/10.4121/uuid:9e625b55-11cf-4de2-8b81-32b5bf04d53d}.

\section{End notes}
\noindent \textbf{Acknowledgments}\\
We gratefully acknowledge Bernard van Heck for fruitful discussions and we thank Alexandra Fursina and Christine Nebel for the growth substrate preparation. This work has been supported by the European Research Council (ERC HELENA 617256 and Synergy), the Dutch Organization for Scientific Research (NWO) and Microsoft Corporation Station Q.\\~\\

\noindent \textbf{Author contributions}\\
F.B., K.Z.\ and S.H.\ fabricated the devices. F.B., S.H.\ and K.Z.\ performed the measurements. F.B.\ analysed the transport data. F.B., S.H.\ and L.P.K.\ discussed the results. S.G.\ and R.L.M.O.h.V.\ carried out the growth and the transfer of the interconnected InSb nanowires under the supervision of E.P.A.M.B.. F.B.\ wrote the manuscript with contributions from S.H.\ and all authors provided critical feedback. S.H.\ and L.P.K.\ supervised the project.\\~\\

\noindent \textbf{Competing Interests}\\
The authors declare no competing interests.
\end{bibunit}


\begin{bibunit}

\renewcommand{\figurename}{\textbf{Supplementary Figure}}
\renewcommand{\thefigure}{\textbf{\arabic{figure}}}

\renewcommand{\tablename}{\textbf{Supplementary Table}}
\renewcommand{\thetable}{\textbf{\arabic{table}}}

\def\bibsection{\section*{Supplementary References}} 

\setcounter{figure}{0}

\clearpage
\widetext
\begin{center}
\LARGE\textbf{Supplementary Information}\\
\bigskip \bigskip \bigskip
\LARGE{Transmission phase read-out of a large quantum dot in a nanowire interferometer}\\
\bigskip \bigskip \bigskip
\LARGE{F. Borsoi \textit{et al.}}
\end{center}

\newpage
\section*{Supplementary Note 1. Transfer of nanowire network devices}
\begin{figure}[hbt!]
\includegraphics{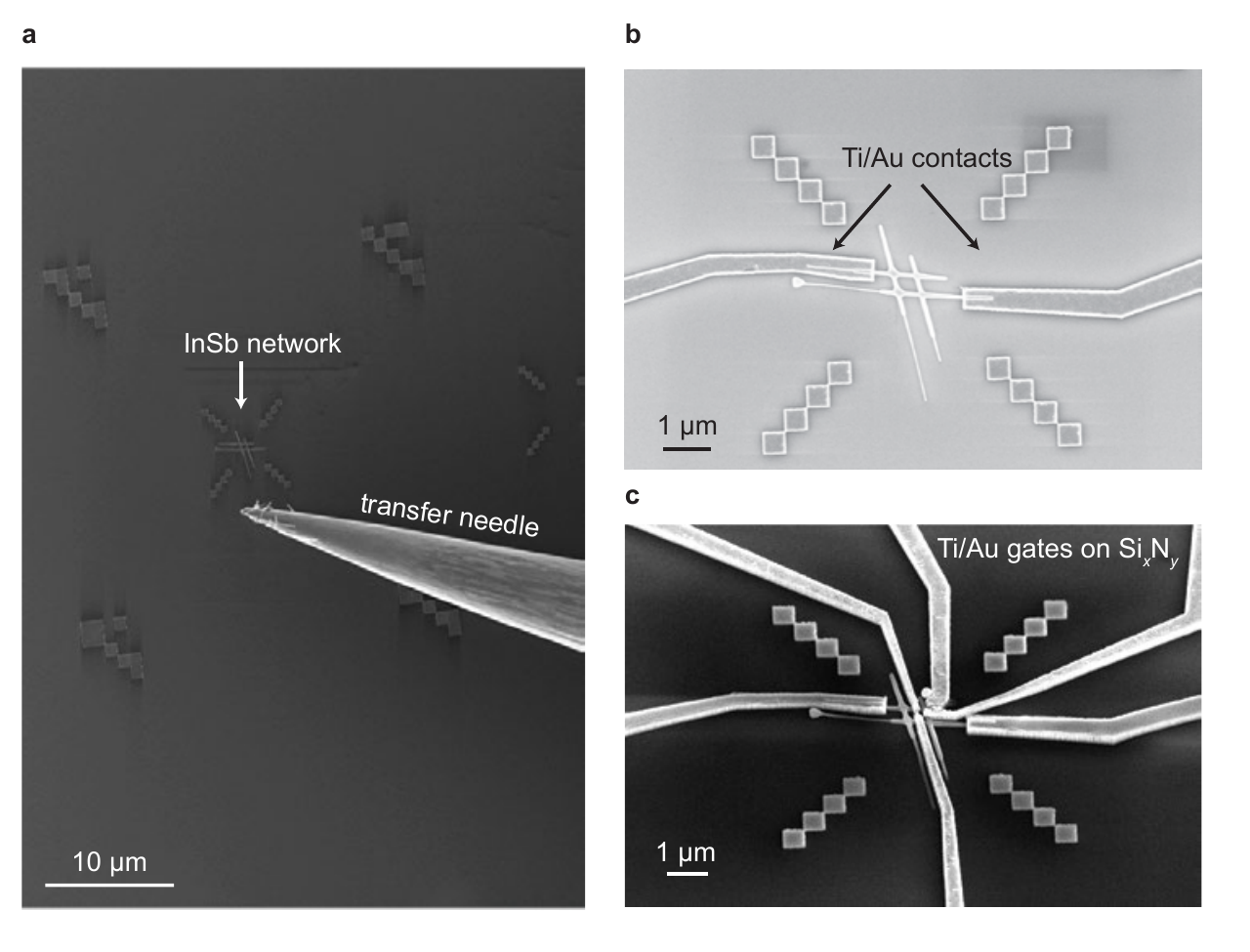}
\caption{Transfer of the InSb nanowire networks. \textbf{a}~Mechanical transfer of a nanowire network with a nanomanipulator in a scanning electron microscope. The device is placed in the vicinity of pre-patterned alignment markers. \textbf{b}, \textbf{c}~Scanning electron micrographs of the device after the fabrication of the Ti/Au contacts and the Ti/Au top gates that are deposited onto a dielectric layer of Si$_x$N$_y$, respectively.}
\label{fig:Supplementary Figure 1}
\end{figure}

\newpage
\section*{Supplementary Note 2. Back-gate dependence}
\begin{figure}[hbt!]
\includegraphics{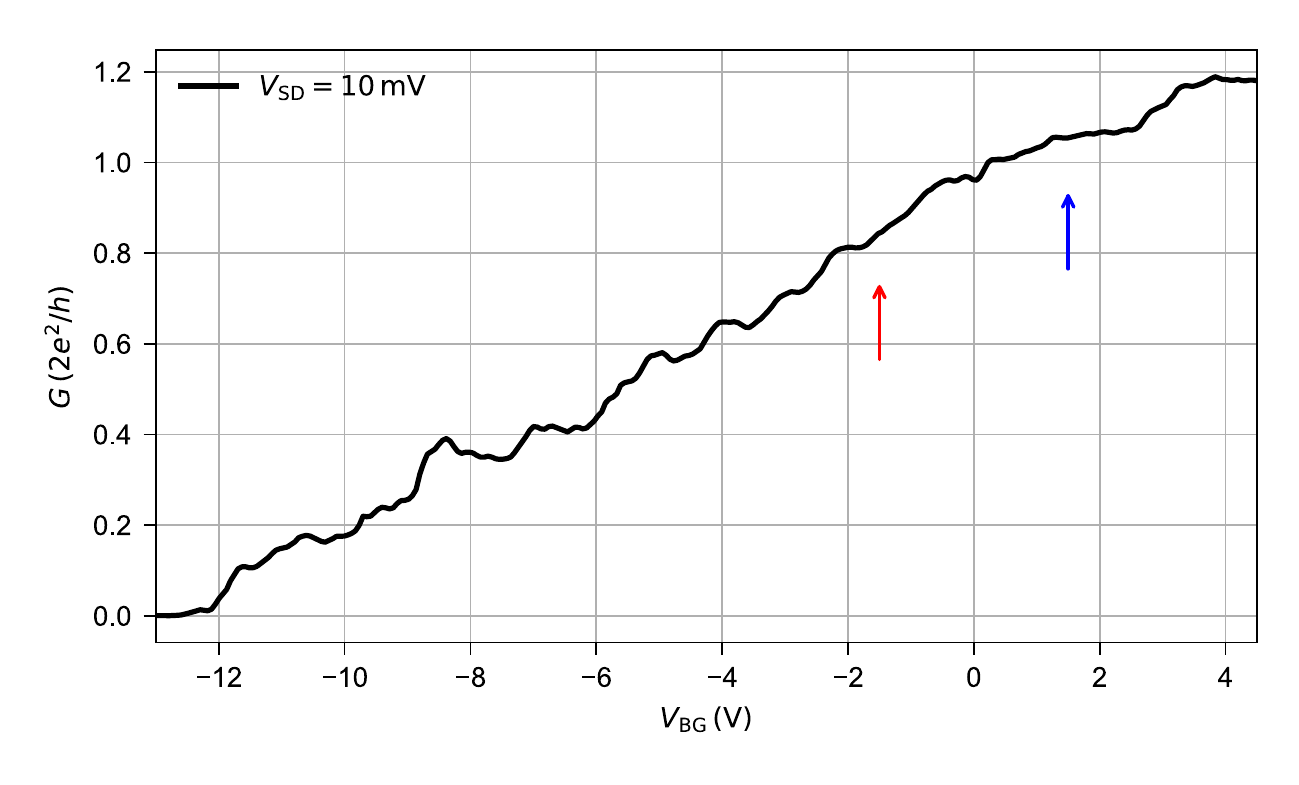}
\caption{$G$ vs.\ $V_{\mathrm{BG}}$ at $10\,$mV bias voltage. The conductance increases linearly with respect to the global back gate in agreement with the Drude model. The blue and red arrows at $1.5\,$V and $-1.5\,$V, respectively, indicate the two working points at which the data discussed in the main text are taken. In the second case, the electron density (roughly proportional to the conductance) is only $\sim 20\%$ lower than in the first case. Therefore, we can conclude that the quantum dot is in both cases in the many-electron regime.}
\label{fig:Supplementary Figure 2}
\end{figure}

\newpage
\section*{Supplementary Note 3. Fourier spectrum of the magneto-conductance oscillations}
\begin{figure}[hbt!]
\includegraphics{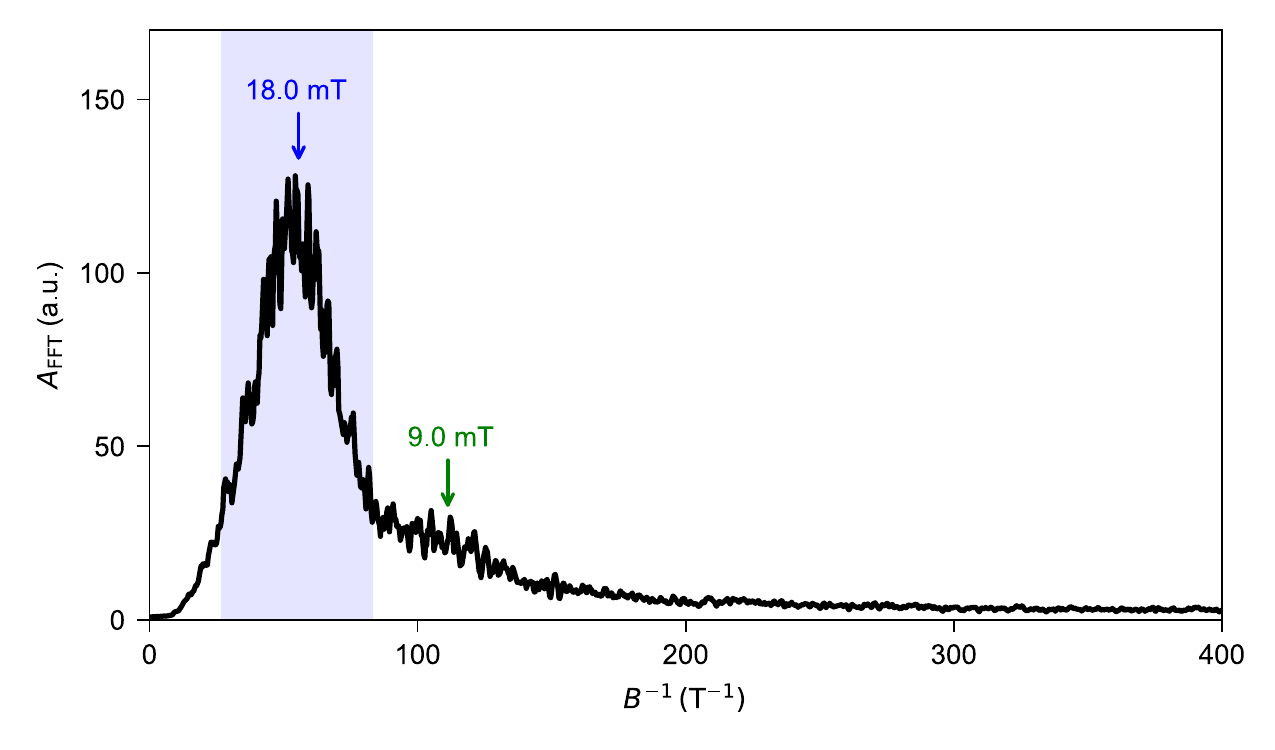}
\caption{Averaged Fourier spectrum of the magneto-conductance oscillations. Eight different magneto-conductance traces are measured and filtered with the Savitzky-Golay algorithm to remove slow oscillations such as universal conductance fluctuations. The Fourier transforms are then calculated and averaged to obtain the trace in the figure. The light blue window identifies the width of the spectrum expected from the area enclosed by the loop. The
peak maximum at $55\,\mathrm{T^{-1}}$ corresponds to the Aharonov-Bohm periodicity of $18\,$mT and perfectly matches the expected value considering the area enclosed by the centre of the interconnected nanowires. A smaller, but discernible peak at $110\,\mathrm{T^{-1}}$ indicates the presence of Altshuler-Aronov-Spivak oscillations.}
\label{fig:Supplementary Figure 3}
\end{figure}
\newpage
\section*{Supplementary Note 4. Additional data in the multi-path Aharonov-Bohm regime}
Two methods were used to study the phase evolution of the magneto-conductance oscillations for the device discussed in the main text. The data shown in Figs.~4 and 5 are taken by sweeping $V_{\mathrm{PG}}$ as the fast axis and $B_{\perp}$ as the slow axis. Oppositely, in Supplementary Figure~\ref{fig:Supplementary Figure 4} we present data taken with $B_{\perp}$ and $V_{\mathrm{PG}}$ being the fast and slow axis, respectively. The data presented here are taken at $ V_{\mathrm{BG}} = 1.5\,$V as in Fig.~4 of the main text. For each value of $V_{\mathrm{PG}}$ (proportional to $E_{\mathrm{QD}}$) the maxima of the AB oscillations are tracked, and their positions in $B_{\perp}$ are converted into the transmission phase $\varphi$ via the AB periodicity and plotted in Supplementary Figure~\ref{fig:Supplementary Figure 4}a. In Supplementary Figure~\ref{fig:Supplementary Figure 4}b, we show the corresponding conductance trace (blue data points) exhibiting CPs. The phase displays a rapid evolution close to the charge degeneracy points. At the two inner CPs, the phase $\varphi$ evolves by $\sim \uppi$, and gradually shifts back to the original value. For the first and fourth CPs, the variation is much smaller than $\uppi$. For all displayed Coulomb valleys, the phase is neither constant, nor does it exhibit a rapid phase lapse. Instead, gradual variations are observed, compatible with the picture discussed in the main text in which multiple trajectories with different enclosed areas in the nanowire interferometer couple to different QD orbitals.

\begin{figure}[hbt!]
\includegraphics{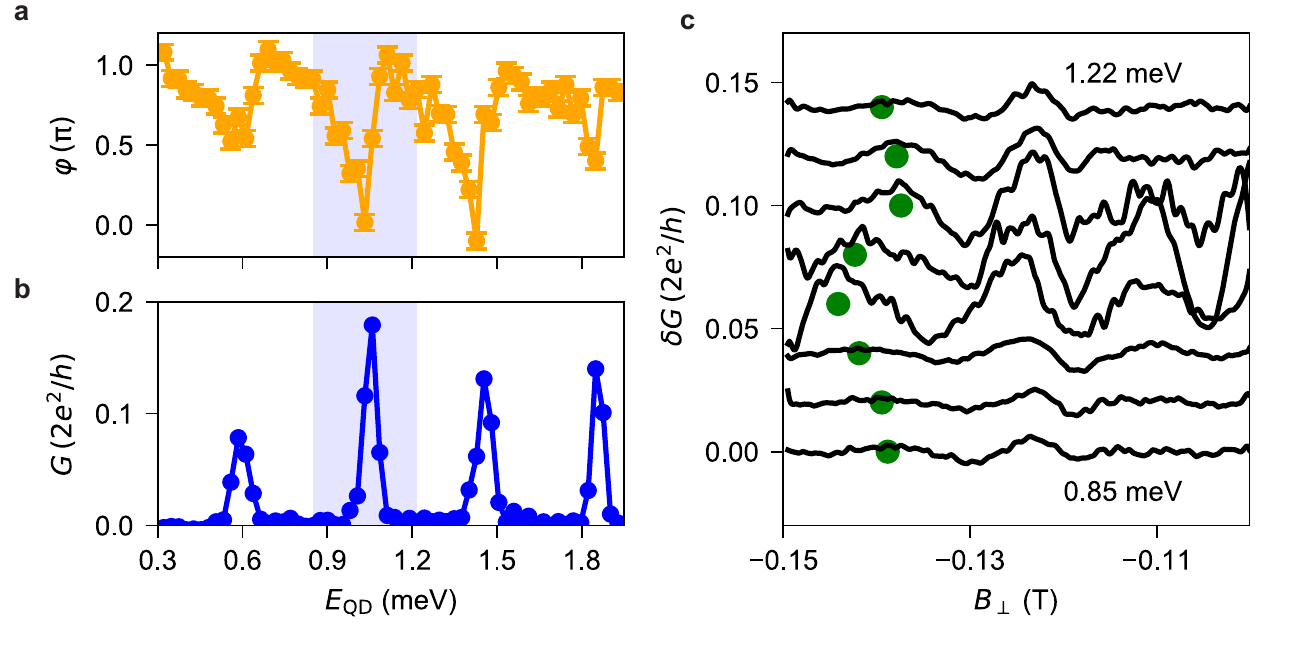}
\caption{Additional data in the multi-path Aharonov-Bohm regime. \textbf{a}~Phase $\varphi$ of the magneto-conductance oscillations as a function of $E_{\mathrm{QD}}$, \textbf{b}~$G$ vs.\ $E_{\mathrm{QD}}$ exhibiting four CPs. \textbf{c}~Aharonov-Bohm oscillations measured in the proximity of the second CP. $\delta G$ is obtained by subtracting a slowly varying background from the magneto-conductance traces. The green circles identify the maxima of the oscillations in the first period. The traces are displaced for clarity.}
\label{fig:Supplementary Figure 4}
\end{figure}

\section*{Supplementary Note 5. Phase variations and fundamental symmetries}
The top panel of Fig.~4d in the main text shows that the transmission phase swings from $\mathrm{0}$ to $\mathrm{\uppi}$ at resonance over an energy range of $\sim \mathit{\Gamma}$. This continuous variation differs significantly from what has been reported in the pioneering experiment of ref.~\onlinecite{YAC1995} where the phase was mysteriously locked to $\mathrm{0}$ and $\mathrm{\uppi}$ with abrupt switches in between. Later on, their finding was understood in term of fundamental symmetries.\\
Time-reversal symmetry imposes, in fact, the two-terminal conductance to be an even function of the magnetic field (the Casimir-Onsager relation: $G(B) = G(-B)$). In an Aharonov-Bohm interferometer, the conductance displays sinusoidal oscillations on top of a background, and the symmetry around zero field imposes on their phase to assume only two values: $\mathrm{0}$ or $\mathrm{\uppi}$. However, in our experiment we find the opposite: the phase variation at resonance is smooth and not abrupt. We associate this behaviour with the fact that time-reversal symmetry does not hold in our context due to the large magnetic fields applied.\\ 

\newpage
\section*{Supplementary Note 6. Fitting results}
\subsection{Main text Fig.~3}
We obtain the dashed lines in Fig.~3a by fitting each peak independently with the single-trajectory model Aharonov-Bohm interferometer. For this and other fits, we use $J=1$ and $ka = 1.57 \sim \uppi/2$ similarly to ref.~\onlinecite{AHA2005}. The phase of $J_\mathrm{L}$ is a fitting parameter that is not relevant and therefore not displayed. We report the other values in Supplementary Table~\ref{tab:Supplementary Table 1} subdivided into the three characteristic regimes in Fig.~3a.

\begin{table}[hbt!]
\centering
\begin{tabular}{|l|l|l|l|}
\cline{2-4}
\multicolumn{1}{c|}{} & \multicolumn{3}{c|}{\textbf{Regime}}\\ \hline
\textbf{Average parameter} & low $t_{\mathrm{ref}}$ (green trace) & medium $t_{\mathrm{ref}}$ (blue trace) & high $t_{\mathrm{ref}}$ (orange trace) \\ \hline
$j_\mathrm{L} \, (\mathrm{\upmu eV})$        & $31\pm1$     & $34\pm6$       & $14\pm8$      \\ \hline
$J_\mathrm{R} \, (\mathrm{\upmu eV})$       & $140\pm6$   & $202\pm17$     & $206\pm26$    \\ \hline
$t_{\mathrm{ref}} $               & $<10^{-3}$  & $0.04\pm0.01$  & $0.28\pm0.02$ \\ \hline
\end{tabular}
\caption{Average and standard deviations of the best-fit parameters of four peaks in the three different regimes in Fig.~3a of the main text.}
\label{tab:Supplementary Table 1}
\end{table}

\subsection{Main text Fig.~4}
We obtain Fig.~4b by fitting the data of Fig.~4a with the single-trajectory model Aharonov-Bohm interferometer with $N = 2$. The fitting procedure is facilitated by fixing two parameters globally (i.e., for the entire measurement): a constant offset to the traces of $0.02 \cdot 2e^2/h$ and $t_{\mathrm{ref}} = 0.07$. The phase of $J_\mathrm{L}$ is an independent parameter that increases with the magnetic field along the y-axis. We display in Supplementary Table~\ref{tab:Supplementary Table 2} the best-fit parameters averaged along the magnetic field axis. Peak numbers are ordered from left to right. 

\begin{table}[hbt!]
\centering
\begin{tabular}{|l|l|l|}
\cline{2-3}
\multicolumn{1}{c|}{} & \multicolumn{2}{c|}{\textbf{Peak number}}\\ \hline
\textbf{Average parameter} & $1$             & $2$             \\ \hline
$j_\mathrm{L} \, (\mathrm{\upmu eV})$           & $192\pm22$       & $182\pm 12$      \\ \hline
$J_\mathrm{R} \, (\mathrm{\upmu eV})$           & $57\pm10$       & $59\pm5$        \\ \hline
$E_{\mathrm{n}} \, (\mathrm{meV}) $    & $0.241\pm0.001$ & $0.618\pm0.011$ \\ \hline
\end{tabular}
\caption{Best-fit parameters averaged along the magnetic field axis of Fig.~4. The errors presented are the standard deviations of the distributions of the parameters.}
\label{tab:Supplementary Table 2}
\end{table}

\subsection{Main text Fig.~5}
In order to obtain Fig.~5b of the main text, we first consider the line-cut of the data at $B_\perp = 0.667\,$T and fit the peaks independently with the single-path interferometer model. We fix some of the parameters to simplify the procedure as $t_{\mathrm{ref}} = 0.05$ and the added offset to the trace at $0.06 \cdot 2e^2/h$. In this procedure, the phase of $J_\mathrm{L}$ is used as a free parameter and not relevant here. The other values are shown in Supplementary Table~\ref{tab:Supplementary Table 3}, where the peak names in Fig.~5a are ordered numerically from left to right.\\
We then extrapolate the trace with the best-fit parameters by varying the phase of $J_\mathrm{L}$ and adjusting the values of $x(n) \in [0.025,\ 0.01,\ 0,\ -0.016]$ to qualitatively reproduce the experimental data.

\begin{table}[hbt!]
\centering
\begin{tabular}{|l|l|l|l|l|}
\cline{2-5}
\multicolumn{1}{c|}{} & \multicolumn{4}{c|}{\textbf{Peak number}}\\ \hline
\textbf{Parameter}                & 1      & 2     & 3      & 4  \\ \hline
\textbf{$j_\mathrm{L} \, (\mathrm{\upmu eV})$}  & $178\pm2$       & $165\pm4$       & $153\pm5$       & $167\pm2$       \\ \hline
\textbf{$J_\mathrm{R} \, (\mathrm{\upmu eV})$} & $33.4\pm0.2$    & $60.9\pm0.1$    & $51.4\pm1.1$    & $37.3\pm0.04$   \\ \hline
\textbf{$E_\mathrm{n} \, (\mathrm{meV})$}     & $2.000\pm0.001$ & $2.293\pm0.002$ & $2.710\pm0.002$ & $3.092\pm0.001$ \\ \hline
\end{tabular}
\caption{Fitting parameters obtained at $B_\perp = 0.667\,$T of Fig.~5a in the main text.}
\label{tab:Supplementary Table 3}
\end{table}

\newpage
\section*{Supplementary Note 7. Cotunnelling Aharonov-Bohm effect in a second device}
\begin{figure}[hbt!]
\includegraphics{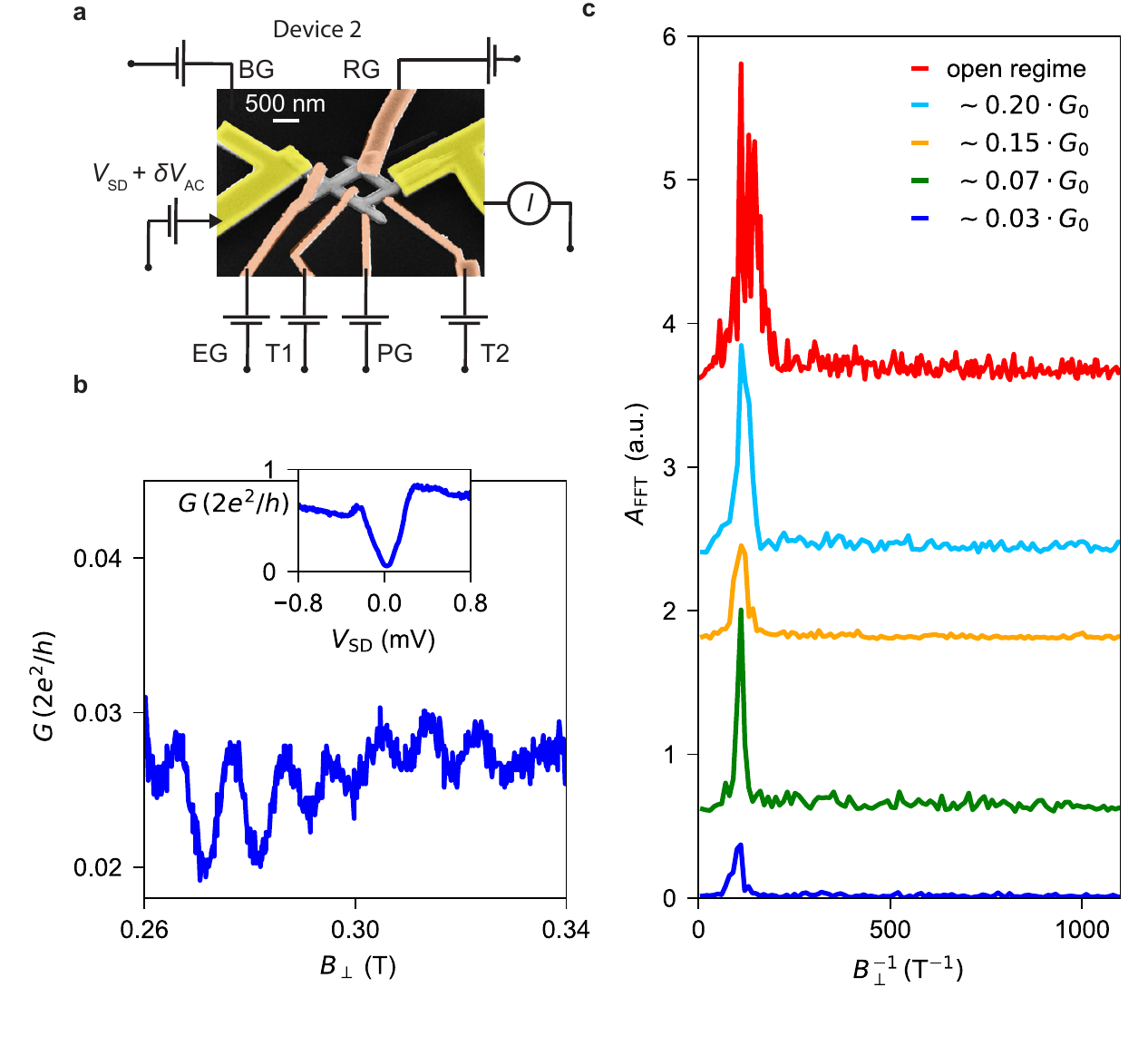}
\caption{Cotunnelling Aharonov-Bohm effect in a second device. \textbf{a}~False-color scanning electron microscopy image of a second device. A quantum dot with a T-shape is formed in the bottom branch of the network by applying negative voltages to the two tunnel gates (T1 and T2). The induced charge is tuned via the plunger gate (PG). We use the top branch of the network as the reference arm of the interferometer, with a transmission tunable by the reference gate (RG). \textbf{b}~Differential conductance $G$ at zero bias voltage $\mathrm{V_{\mathrm{SD}}}$ as a function of the perpendicular field $B_{\perp}$ in the cotunnelling regime manifesting AB oscillations with a period of $\sim 9 - 11\,$mT. Inset: $G$ vs.\ $V_{\mathrm{SD}}$ taken at $B_{\perp} = 0.34\,$T. \textbf{c}~Magnitude of the Fast Fourier Transforms for different tunnel gate settings. The legend indicates the average value of the cotunnelling conductance of the raw magneto-conductance traces.}
\label{fig:Supplementary Figure 5}
\end{figure}

Here, we report additional measurements of Aharonov-Bohm interference in the cotunnelling regime for a second device. The electron density in this device (Supplementary Figure~\ref{fig:Supplementary Figure 5}a) is tunable by voltages on a global back gate and several top gates. The top dielectric used here is $\mathrm{Al_2 O_3}$ grown via atomic layer deposition. A T-shaped quantum dot is formed in the bottom branch of the interferometer and has a charging energy of $E_{\mathrm{c}} \sim 0.5\,$meV, with the tunnel coupling to the leads adjustable using the tunnel gates T1 and T2. Although the conductance around $V_{\mathrm{SD}} = 0$ is strongly suppressed owing to the Coulomb blockade (inset of Supplementary Figure~\ref{fig:Supplementary Figure 5}b), the magneto-conductance exhibits clear Aharonov-Bohm oscillations with a periodicity of $9-11\,$mT (Supplementary Figure~\ref{fig:Supplementary Figure 5}b, main panel). The oscillation amplitude is sizable despite the cotunnelling conductance being as low as $ \sim 0.025 \cdot 2e^2/h$. This value corresponds to a typical dwell time in the QD in the order of $5-10\,$ps. The magnitude of the Fast Fourier Transform (FFT) of these oscillations is displayed in Supplementary Figure~\ref{fig:Supplementary Figure 5}c, together with the FFTs of the AB oscillations at gradually stronger cotunnelling conductance, until the Coulomb blockade is fully quenched (red trace corresponding to the 'open regime' with $G\sim 2e^2/h$). We observe here that the width of the FFT peak of the AB signal increases when quenching the Coulomb blockade by making the barriers more transparent (in particular, see the difference between the red and the other traces). This fact might suggest that, when the quantum dot is defined, fewer trajectories play a role in the transport through the interferometer, reducing the spread of the enclosed area.

\newpage
\begin{figure}[hbt!]
\includegraphics{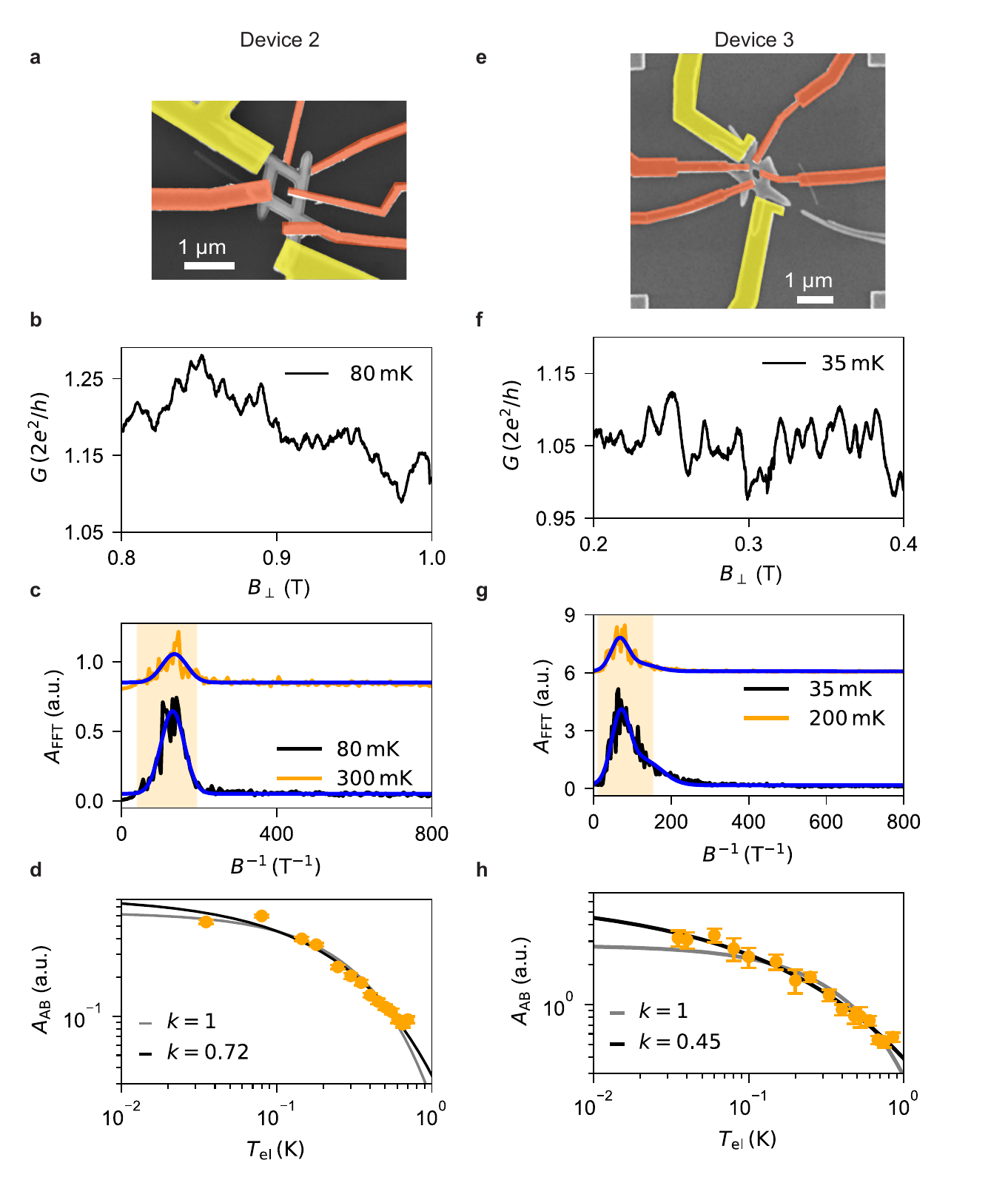}
\caption{Aharonov-Bohm temperature dependence of the second and third device. \textbf{a}, \textbf{e}~Scanning electron micrographs of devices 2 and 3, respectively. \textbf{b}, \textbf{f}~Typical magneto-conductance traces for the two devices exhibiting oscillations superimposed on a slowly varying background. \textbf{c}, \textbf{g}~Average of several Fast Fourier Transforms of oscillations at two different temperatures fitted with a Gaussian in \textbf{c} and a sum of two Gaussians in \textbf{g}. The yellow shaded area denotes the expected range of the first harmonic (Aharonov-Bohm) according to the sizes of the loops. In \textbf{d}, \textbf{h} the temperature dependence of the AB amplitudes is presented for both the devices (orange points). Black and grey traces are the best fits of the data.}
\label{fig:Supplementary Figure 6}
\end{figure}

\section*{Supplementary Note 8. Temperature dependence of the Aharonov-Bohm effect}
In Supplementary Figures~\ref{fig:Supplementary Figure 6}a and \ref{fig:Supplementary Figure 6}e, we show the scanning electron micrographs of two nanowire loops (second and third devices) used to study the temperature dependence of the Aharonov-Bohm oscillations. In Supplementary Figures~\ref{fig:Supplementary Figure 6}b and \ref{fig:Supplementary Figure 6}f, we plot two typical magneto-conductance traces of each device. They both exhibit AB oscillations superimposed on a slowly varying background. In the data analysis, we subtract this background and calculate the Fast Fourier Transform. The amplitudes $A_{\mathrm{FFT}}$ shown in Figs.~\ref{fig:Supplementary Figure 6}c and \ref{fig:Supplementary Figure 6}g are the averaged Fast Fourier Transforms of several scans taken in the same magnetic field window. The peaks in the spectra are fitted with a Gaussian obtaining the power spectrum of the AB oscillations $A_{\mathrm{AB}}$. For the third device, we find that the sum of two Gaussian curves better describes the broad FFT peak due to the presence of the second harmonic. In panels d and h, we demonstrate the decrease of $A_{\mathrm{AB}}$ as a function of temperature. The amplitudes of each harmonic (with index $n$) are expected to decrease as $\exp(-n\cdot L/l_\mathrm{\phi})$, with $l_\mathrm{\phi}$ being the phase coherence length and $L$ being the effectively travelled path length~\cite{IHN2010}. In ballistic systems, we have $l_\mathrm{\phi} \propto T^{-1}$, while $l_\mathrm{\phi} \propto T^{-1/2}$ in the diffusive regime~\cite{LUD2004}.
We fitted the experimental decay of the Aharonov-Bohm amplitude with the function $A\cdot \exp(-\alpha T^k)$, with free parameters $A$, $\alpha$, $k$, and the Aharonov-Bohm phase coherence length defined as $l_{\mathrm{\phi}} = (L/\alpha) T^{-k}$. The best-fit curves are shown in black, and correspond to $k_{\mathrm{best}} = 0.72$ and $k_{\mathrm{best}} = 0.45$ for the second and third devices, respectively. For comparison, we also show the best-fit curves obtained for $k = 1$. The values of the fitted exponents $k$ reflect the nature of the transport in the semiconducting loops, which is in the crossover between completely ballistic and diffusive. In fact, the typically travelled length is $\sim 1\,\upmu$m, which is only a few times larger than the estimated mean free path of $\sim 0.1 - 0.3\,\upmu$m~\cite{VAN2015, GUL2015}. We expect that at higher temperatures and for the longer loops (cf.\ device 2) the diffusive model with $k = 0.5$ captures the experimental trend better than the ballistic one, but the experimental data do not provide conclusive evidence.\\
From our analysis, we derive $l_{\mathrm{\phi}} (35\,\mathrm{mK}) = \frac{L}{\alpha} T^{-k} \sim 8\,\upmu$m for the second device, and $l_{\phi} (35\,\mathrm{mK}) \sim 2.5\,\upmu$m for the third loop. It can be argued that these values are probably underestimating the actual coherence length since the amplitude of the AB oscillations (more generally all odd harmonics) are additionally suppressed by the energy averaging effect.

\end{bibunit}


\begin{thebibliography}{10}
\expandafter\ifx\csname url\endcsname\relax
  \def\url#1{\texttt{#1}}\fi
\expandafter\ifx\csname urlprefix\endcsname\relax\def\urlprefix{URL }\fi
\providecommand{\bibinfo}[2]{#2}
\providecommand{\eprint}[2][]{\url{#2}}

\bibitem{WEB1985}
\bibinfo{author}{Webb, R.~A.}, \bibinfo{author}{Washburn, S.},
  \bibinfo{author}{Umbach, C.~P.} \& \bibinfo{author}{Laibowitz, R.~B.}
\newblock \bibinfo{title}{Observation of h/e \mbox{A}haronov-\mbox{B}ohm
  oscillations in normal-metal rings}.
\newblock \emph{\bibinfo{journal}{Phys. Rev. Lett.}}
  \textbf{\bibinfo{volume}{54}}, \bibinfo{pages}{2696--2699}
  (\bibinfo{year}{1985}).

\bibitem{BAC1999}
\bibinfo{author}{Bachtold, A.} \emph{et~al.}
\newblock \bibinfo{title}{Aharonov-\mbox{B}ohm oscillations in carbon
  nanotubes}.
\newblock \emph{\bibinfo{journal}{Nature}} \textbf{\bibinfo{volume}{397}},
  \bibinfo{pages}{673--675} (\bibinfo{year}{1999}).

\bibitem{RUS2008}
\bibinfo{author}{Russo, S.} \emph{et~al.}
\newblock \bibinfo{title}{Observation of \mbox{A}haronov-\mbox{B}ohm
  conductance oscillations in a graphene ring}.
\newblock \emph{\bibinfo{journal}{Phys. Rev. B}} \textbf{\bibinfo{volume}{77}},
  \bibinfo{pages}{085413} (\bibinfo{year}{2008}).

\bibitem{LEE1998}
\bibinfo{author}{Lee, H.-W.}
\newblock \bibinfo{title}{Generic transmission zeros and in-phase resonances in
  time-reversal symmetric single channel transport}.
\newblock \emph{\bibinfo{journal}{Phys. Rev. Lett.}}
  \textbf{\bibinfo{volume}{82}}, \bibinfo{pages}{2358--2361} (\bibinfo{year}{1998}).

\bibitem{ORE2007}
\bibinfo{author}{Oreg, Y.}
\newblock \bibinfo{title}{Universal phase lapses in a noninteracting model}.
\newblock \emph{\bibinfo{journal}{New J. Phys.}} \textbf{\bibinfo{volume}{9}},
  \bibinfo{pages}{122},
  \bibinfo{url}{\url{https://doi.org/10.1088/1367-2630/9/5/122}}
  (\bibinfo{year}{2007}).

\bibitem{SIL200}
\bibinfo{author}{Silvestrov, P.~G.} \& \bibinfo{author}{Imry, Y.}
\newblock \bibinfo{title}{Towards an explanation of the mesoscopic double-slit
  experiment: A new model for charging of a quantum dot}.
\newblock \emph{\bibinfo{journal}{Phys. Rev. Lett.}}
  \textbf{\bibinfo{volume}{85}}, \bibinfo{pages}{2565--2568}
  (\bibinfo{year}{2000}).

\bibitem{VAN2006}
\bibinfo{author}{van Dam, J.~A.}, \bibinfo{author}{Nazarov, Y.~V.},
  \bibinfo{author}{Bakkers, E. P. A.~M.}, \bibinfo{author}{De~Franceschi, S.}
  \& \bibinfo{author}{Kouwenhoven, L.~P.}
\newblock \bibinfo{title}{Supercurrent reversal in quantum dots}.
\newblock \emph{\bibinfo{journal}{Nature}} \textbf{\bibinfo{volume}{442}},
  \bibinfo{pages}{667--670} (\bibinfo{year}{2006}).

\bibitem{KIT2001}
\bibinfo{author}{Kitaev, A.~Y.}
\newblock \bibinfo{title}{Unpaired \mbox{M}ajorana fermions in quantum wires}.
\newblock \emph{\bibinfo{journal}{Phys.-Uspekhi}}
  \textbf{\bibinfo{volume}{44}}, \bibinfo{pages}{131--136}
  (\bibinfo{year}{2001}).

\bibitem{LUT2010}
\bibinfo{author}{Lutchyn, R.~M.}, \bibinfo{author}{Sau, J.~D.} \&
  \bibinfo{author}{Das~Sarma, S.}
\newblock \bibinfo{title}{Majorana fermions and a topological phase transition
  in semiconductor-superconductor heterostructures}.
\newblock \emph{\bibinfo{journal}{Phys. Rev. Lett.}}
  \textbf{\bibinfo{volume}{105}}, \bibinfo{pages}{077001}
  (\bibinfo{year}{2010}).

\bibitem{ORE2010}
\bibinfo{author}{Oreg, Y.}, \bibinfo{author}{Refael, G.} \&
  \bibinfo{author}{von Oppen, F.}
\newblock \bibinfo{title}{Helical liquids and \mbox{M}ajorana bound states in
  quantum wires}.
\newblock \emph{\bibinfo{journal}{Phys. Rev. Lett.}}
  \textbf{\bibinfo{volume}{105}}, \bibinfo{pages}{177002}
  (\bibinfo{year}{2010}).

\bibitem{ALI2012}
\bibinfo{author}{Alicea, J.}
\newblock \bibinfo{title}{New directions in the pursuit of \mbox{M}ajorana
  fermions in solid state systems}.
\newblock \emph{\bibinfo{journal}{Rep. Prog. Phys.}}
  \textbf{\bibinfo{volume}{75}}, \bibinfo{pages}{076501}
  (\bibinfo{year}{2012}).

\bibitem{VIJ2016}
\bibinfo{author}{Vijay, S.} \& \bibinfo{author}{Fu, L.}
\newblock \bibinfo{title}{Teleportation-based quantum information processing
  with \mbox{M}ajorana zero modes}.
\newblock \emph{\bibinfo{journal}{Phys. Rev. B}} \textbf{\bibinfo{volume}{94}},
  \bibinfo{pages}{235446} (\bibinfo{year}{2016}).

\bibitem{PLU2017}
\bibinfo{author}{Plugge, S.}, \bibinfo{author}{Rasmussen, A.},
  \bibinfo{author}{Egger, R.} \& \bibinfo{author}{Flensberg, K.}
\newblock \bibinfo{title}{Majorana box qubits}.
\newblock \emph{\bibinfo{journal}{New J. Phys.}} \textbf{\bibinfo{volume}{19}},
  \bibinfo{pages}{012001},
  \bibinfo{url}{\url{https://doi.org/10.1088/1367-2630/aa54e1}}
  (\bibinfo{year}{2017}).

\bibitem{KAR2017}
\bibinfo{author}{Karzig, T.} \emph{et~al.}
\newblock \bibinfo{title}{Scalable designs for
  quasiparticle-poisoning-protected topological quantum computation with
  \mbox{M}ajorana zero modes}.
\newblock \emph{\bibinfo{journal}{Phys. Rev. B}} \textbf{\bibinfo{volume}{95}},
  \bibinfo{pages}{235305} (\bibinfo{year}{2017}).

\bibitem{FU2010}
\bibinfo{author}{Fu, L.}
\newblock \bibinfo{title}{Electron teleportation via \mbox{M}ajorana bound
  states in a mesoscopic superconductor}.
\newblock \emph{\bibinfo{journal}{Phys. Rev. Lett.}}
  \textbf{\bibinfo{volume}{104}}, \bibinfo{pages}{056402}
  (\bibinfo{year}{2010}).

\bibitem{DRU2018}
\bibinfo{author}{Drukier, C.}, \bibinfo{author}{Zirnstein, H.-G.},
  \bibinfo{author}{Rosenow, B.}, \bibinfo{author}{Stern, A.} \&
  \bibinfo{author}{Oreg, Y.}
\newblock \bibinfo{title}{Evolution of the transmission phase through a
  \mbox{C}oulomb-blockaded \mbox{M}ajorana wire}.
\newblock \emph{\bibinfo{journal}{Phys. Rev. B}} \textbf{\bibinfo{volume}{98}},
  \bibinfo{pages}{161401} (\bibinfo{year}{2018}).

\bibitem{HEL2018}
\bibinfo{author}{Hell, M.}, \bibinfo{author}{Flensberg, K.} \&
  \bibinfo{author}{Leijnse, M.}
\newblock \bibinfo{title}{Distinguishing \mbox{M}ajorana bound states from
  localized \mbox{A}ndreev bound states by interferometry}.
\newblock \emph{\bibinfo{journal}{Phys. Rev. B}} \textbf{\bibinfo{volume}{97}},
  \bibinfo{pages}{161401} (\bibinfo{year}{2018}).

\bibitem{AVI2005}
\bibinfo{author}{Avinun-Kalish, M.}, \bibinfo{author}{Heiblum, M.},
  \bibinfo{author}{Zarchin, O.}, \bibinfo{author}{Mahalu, D.} \&
  \bibinfo{author}{Umansky, V.}
\newblock \bibinfo{title}{Crossover from ‘mesoscopic’ to ‘universal’
  phase for electron transmission in quantum dots}.
\newblock \emph{\bibinfo{journal}{Nature}} \textbf{\bibinfo{volume}{436}}, 
  \bibinfo{pages}{529--533} (\bibinfo{year}{2005}).

\bibitem{YAC1995}
\bibinfo{author}{Yacoby, A.}, \bibinfo{author}{Heiblum, M.},
  \bibinfo{author}{Mahalu, D.} \& \bibinfo{author}{Shtrikman, H.}
\newblock \bibinfo{title}{Coherence and phase sensitive measurements in a
  quantum dot}.
\newblock \emph{\bibinfo{journal}{Phys. Rev. Lett.}}
  \textbf{\bibinfo{volume}{74}}, \bibinfo{pages}{4047--4050}
  (\bibinfo{year}{1995}).

\bibitem{SCH1997}
\bibinfo{author}{Schuster, R.} \emph{et~al.}
\newblock \bibinfo{title}{Phase measurement in a quantum dot via a double-slit
  interference experiment}.
\newblock \emph{\bibinfo{journal}{Nature}} \textbf{\bibinfo{volume}{385}},
  \bibinfo{pages}{417--420} (\bibinfo{year}{1997}).

\bibitem{KAR2007}
\bibinfo{author}{Karrasch, C.} \emph{et~al.}
\newblock \bibinfo{title}{Mesoscopic to universal crossover of the transmission
  phase of multilevel quantum dots}.
\newblock \emph{\bibinfo{journal}{Phys. Rev. Lett.}}
  \textbf{\bibinfo{volume}{98}}, \bibinfo{pages}{186802}
  (\bibinfo{year}{2007}).

\bibitem{EDL2017}
\bibinfo{author}{Edlbauer, H.} \emph{et~al.}
\newblock \bibinfo{title}{Non-universal transmission phase behaviour of a large
  quantum dot}.
\newblock \emph{\bibinfo{journal}{Nat. Commun.}} \textbf{\bibinfo{volume}{8}}, 
  \bibinfo{pages}{1710}, 
  \bibinfo{url}{\url{https://doi.org/10.1038/s41467-017-01685-z}}
  (\bibinfo{year}{2017}).

\bibitem{CAR2014}
\bibinfo{author}{Car, D.}, \bibinfo{author}{Wang, J.},
  \bibinfo{author}{Verheijen, M.~A.}, \bibinfo{author}{Bakkers, E. P. A.~M.} \&
  \bibinfo{author}{Plissard, S.~R.}
\newblock \bibinfo{title}{Rationally designed single-crystalline nanowire
  networks}.
\newblock \emph{\bibinfo{journal}{Adv. Mater.}} \textbf{\bibinfo{volume}{26}},
  \bibinfo{pages}{4875–4879} (\bibinfo{year}{2014}).

\bibitem{GAZ2017}
\bibinfo{author}{Gazibegovic, S.} \emph{et~al.}
\newblock \bibinfo{title}{Epitaxy of advanced nanowire quantum devices}.
\newblock \emph{\bibinfo{journal}{Nature}} \textbf{\bibinfo{volume}{584}}, 
  \bibinfo{pages}{434--438}
  (\bibinfo{year}{2017}).

\bibitem{KOU2001}
\bibinfo{author}{Kouwenhoven, L.~P.}, \bibinfo{author}{Austing, D.~G.} \&
  \bibinfo{author}{Tarucha, S.}
\newblock \bibinfo{title}{Few-electron quantum dots}.
\newblock \emph{\bibinfo{journal}{Rep. Prog. Phys.}}
  \textbf{\bibinfo{volume}{64}}, \bibinfo{pages}{701--736}
  (\bibinfo{year}{2001}).

\bibitem{IHN2010}
\bibinfo{author}{Ihn, T.}
\newblock \emph{\bibinfo{title}{Semiconductor Nanostructures}}
  (\bibinfo{publisher}{Oxford University Press}, \bibinfo{year}{2010}).

\bibitem{ALB2016}
\bibinfo{author}{Albrecht, S.~M.} \emph{et~al.}
\newblock \bibinfo{title}{Exponential protection of zero modes in
  \mbox{M}ajorana islands}.
\newblock \emph{\bibinfo{journal}{Nature}} \textbf{\bibinfo{volume}{531}},
  \bibinfo{pages}{206--209} (\bibinfo{year}{2016}).

\bibitem{VAN2018}
\bibinfo{author}{van Veen, J.} \emph{et~al.}
\newblock \bibinfo{title}{Magnetic-field-dependent quasiparticle dynamics of
  nanowire single-\mbox{C}ooper-pair transistors}.
\newblock \emph{\bibinfo{journal}{Phys. Rev. B}} \textbf{\bibinfo{volume}{98}},
  \bibinfo{pages}{174502} (\bibinfo{year}{2018}).

\bibitem{SHE2018}
\bibinfo{author}{Shen, J.} \emph{et~al.}
\newblock \bibinfo{title}{Parity transitions in the superconducting ground
  state of hybrid \mbox{InSb}-\mbox{Al} \mbox{C}oulomb islands}.
\newblock \emph{\bibinfo{journal}{Nat. Commun.}} \textbf{\bibinfo{volume}{9}},
  \bibinfo{pages}{4801}, 
  \bibinfo{url}{\url{https://doi.org/10.1038/s41467-018-07279-7}}
  (\bibinfo{year}{2018}).

\bibitem{PLI2013}
\bibinfo{author}{Plissard, S.~R.} \emph{et~al.}
\newblock \bibinfo{title}{Formation and electronic properties of \mbox{InSb}
  nanocrosses}.
\newblock \emph{\bibinfo{journal}{Nat. Nanotechnol.}}
  \textbf{\bibinfo{volume}{8}}, \bibinfo{pages}{859--864} (\bibinfo{year}{2013}).

\bibitem{CER1973}
\bibinfo{author}{Cerdeira, F.}, \bibinfo{author}{Fjeldly, T.~A.} \&
  \bibinfo{author}{Cardona, M.}
\newblock \bibinfo{title}{Effect of free carriers on zone-center vibrational
  modes in heavily doped $p$-type \mbox{Si}}.
\newblock \emph{\bibinfo{journal}{Phys. Rev. B}} \textbf{\bibinfo{volume}{8}},
  \bibinfo{pages}{4734--4745} (\bibinfo{year}{1973}).

\bibitem{GUP2003}
\bibinfo{author}{Gupta, R.}, \bibinfo{author}{Xiong, Q.}, \bibinfo{author}{Adu,
  C.~K.}, \bibinfo{author}{Kim, U.~J.} \& \bibinfo{author}{Eklund, P.~C.}
\newblock \bibinfo{title}{Laser-induced \mbox{F}ano resonance scattering in
  silicon nanowires}.
\newblock \emph{\bibinfo{journal}{Nano Lett.}} \textbf{\bibinfo{volume}{3}},
  \bibinfo{pages}{627--631} (\bibinfo{year}{2003}).

\bibitem{FAI1997}
\bibinfo{author}{Faist, J.}, \bibinfo{author}{Capasso, F.},
  \bibinfo{author}{Sirtori, C.}, \bibinfo{author}{West, K.~W.} \&
  \bibinfo{author}{Pfeiffer, L.~N.}
\newblock \bibinfo{title}{Controlling the sign of quantum interference by
  tunnelling from quantum wells}.
\newblock \emph{\bibinfo{journal}{Nature}} \textbf{\bibinfo{volume}{390}},
  \bibinfo{pages}{589--591} (\bibinfo{year}{1997}).

\bibitem{SCHM1997}
\bibinfo{author}{Schmidt, H.}, \bibinfo{author}{Campman, K.~L.},
  \bibinfo{author}{Gossard, A.~C.} \& \bibinfo{author}{Imamo\v{g}lu, A.}
\newblock \bibinfo{title}{Tunneling induced transparency: \mbox{F}ano
  interference in intersubband transitions}.
\newblock \emph{\bibinfo{journal}{Appl. Phys. Lett.}}
  \textbf{\bibinfo{volume}{70}}, \bibinfo{pages}{3455--3457}
  (\bibinfo{year}{1997}).

\bibitem{GOR2000}
\bibinfo{author}{G\"ores, J.} \emph{et~al.}
\newblock \bibinfo{title}{\mbox{F}ano resonances in electronic transport
  through a single-electron transistor}.
\newblock \emph{\bibinfo{journal}{Phys. Rev. B}} \textbf{\bibinfo{volume}{62}},
  \bibinfo{pages}{2188--2194} (\bibinfo{year}{2000}).

\bibitem{KOB2002}
\bibinfo{author}{Kobayashi, K.}, \bibinfo{author}{Aikawa, H.},
  \bibinfo{author}{Katsumoto, S.} \& \bibinfo{author}{Iye, Y.}
\newblock \bibinfo{title}{Tuning of the \mbox{F}ano effect through a quantum
  dot in an \mbox{A}haronov-\mbox{B}ohm interferometer}.
\newblock \emph{\bibinfo{journal}{Phys. Rev. Lett.}}
  \textbf{\bibinfo{volume}{88}}, \bibinfo{pages}{256806}
  (\bibinfo{year}{2002}).

\bibitem{AHA2006}
\bibinfo{author}{Aharony, A.} \emph{et~al.}
\newblock \bibinfo{title}{Breakdown of phase rigidity and variations of the
  \mbox{F}ano effect in closed \mbox{A}haronov-\mbox{B}ohm interferometers}.
\newblock \emph{\bibinfo{journal}{Phys. Rev. B}} \textbf{\bibinfo{volume}{73}},
  \bibinfo{pages}{195329} (\bibinfo{year}{2006}).

\bibitem{HUA2015}
\bibinfo{author}{Huang, L.}, \bibinfo{author}{Lai, Y.-C.},
  \bibinfo{author}{Luo, H.-G.} \& \bibinfo{author}{Grebogi, C.}
\newblock \bibinfo{title}{Universal formalism of \mbox{F}ano resonance}.
\newblock \emph{\bibinfo{journal}{AIP Adv.}} \textbf{\bibinfo{volume}{5}}, 
  \bibinfo{pages}{017137} 
  (\bibinfo{year}{2015}).

\bibitem{RYU1998}
\bibinfo{author}{Ryu, C.-M.} \& \bibinfo{author}{Cho, S.~Y.}
\newblock \bibinfo{title}{Phase evolution of the transmission coefficient in an
  \mbox{A}haronov-\mbox{B}ohm ring with \mbox{F}ano resonance}.
\newblock \emph{\bibinfo{journal}{Phys. Rev. B}} \textbf{\bibinfo{volume}{58}},
  \bibinfo{pages}{3572--3575} (\bibinfo{year}{1998}).

\bibitem{KOB2003}
\bibinfo{author}{Kobayashi, K.}, \bibinfo{author}{Aikawa, H.},
  \bibinfo{author}{Katsumoto, S.} \& \bibinfo{author}{Iye, Y.}
\newblock \bibinfo{title}{Mesoscopic \mbox{F}ano effect in a quantum dot
  embedded in an \mbox{A}haronov-\mbox{B}ohm ring}.
\newblock \emph{\bibinfo{journal}{Phys. Rev. B}} \textbf{\bibinfo{volume}{68}},
  \bibinfo{pages}{235304} (\bibinfo{year}{2003}).

\bibitem{KAT2004}
\bibinfo{author}{Katsumoto, S.}, \bibinfo{author}{Kobayashi, K.},
  \bibinfo{author}{Aikawa, H.}, \bibinfo{author}{Sano, A.} \&
  \bibinfo{author}{Iye, Y.}
\newblock \bibinfo{title}{Quantum coherence in quantum dot -
  \mbox{A}haronov-\mbox{B}ohm ring hybrid systems}.
\newblock \emph{\bibinfo{journal}{Superlattices Microstruct.}}
  \textbf{\bibinfo{volume}{34}}, \bibinfo{pages}{151--157}
  (\bibinfo{year}{2003}).

\bibitem{WHI2019}
\bibinfo{author}{Whiticar, A.~M.} \emph{et~al.}
\newblock \bibinfo{title}{Interferometry and coherent single-electron transport
  through hybrid superconductor-semiconductor \mbox{C}oulomb islands.}
  \bibinfo{note}{Preprint at \url{https://arxiv.org/abs/1902.07085} (2019)}.

\bibitem{PRA2019}
\bibinfo{author}{Prada, E.} \emph{et~al.}
\newblock \bibinfo{title}{From \mbox{A}ndreev to \mbox{M}ajorana bound states
  in hybrid superconductor-semiconductor nanowires.} \bibinfo{note}{Preprint at
  \url{https://arxiv.org/abs/1911.04512} (2019)}.

\end{thebibliography}

\begin{thebibliography}{1}
\expandafter\ifx\csname url\endcsname\relax
  \def\url#1{\texttt{#1}}\fi
\expandafter\ifx\csname urlprefix\endcsname\relax\def\urlprefix{URL }\fi
\providecommand{\bibinfo}[2]{#2}
\providecommand{\eprint}[2][]{\url{#2}}

\bibitem{YAC1995}
\bibinfo{author}{Yacoby, A.}, \bibinfo{author}{Heiblum, M.},
  \bibinfo{author}{Mahalu, D.} \& \bibinfo{author}{Shtrikman, H.}
\newblock \bibinfo{title}{Coherence and phase sensitive measurements in a
  quantum dot}.
\newblock \emph{\bibinfo{journal}{Phys. Rev. Lett.}}
  \textbf{\bibinfo{volume}{74}}, \bibinfo{pages}{4047--4050}
  (\bibinfo{year}{1995}).

\bibitem{AHA2005}
\bibinfo{author}{Aharony, A.}, \bibinfo{author}{Entin-Wohlman, O.} \&
  \bibinfo{author}{Imry, Y.}
\newblock \bibinfo{title}{Phase measurements in open and closed
  \mbox{A}haronov-\mbox{B}ohm interferometers}.
\newblock \emph{\bibinfo{journal}{Phys. E}} \textbf{\bibinfo{volume}{29}},
  \bibinfo{pages}{283 -- 288} (\bibinfo{year}{2005}).

\bibitem{IHN2010}
\bibinfo{author}{Ihn, T.}
\newblock \emph{\bibinfo{title}{Semiconductor Nanostructures}}
  (\bibinfo{publisher}{Oxford University Press}, \bibinfo{year}{2010}).

\bibitem{LUD2004}
\bibinfo{author}{Ludwig, T.} \& \bibinfo{author}{Mirlin, A.~D.}
\newblock \bibinfo{title}{Interaction-induced dephasing of
  \mbox{A}haronov-\mbox{B}ohm oscillations}.
\newblock \emph{\bibinfo{journal}{Phys. Rev. B}} \textbf{\bibinfo{volume}{69}},
  \bibinfo{pages}{193306} (\bibinfo{year}{2004}).

\bibitem{VAN2015}
\bibinfo{author}{van Weperen, I.} \emph{et~al.}
\newblock \bibinfo{title}{Spin-orbit interaction in \mbox{InSb} nanowires}.
\newblock \emph{\bibinfo{journal}{Phys. Rev. B}} \textbf{\bibinfo{volume}{91}},
  \bibinfo{pages}{201413} (\bibinfo{year}{2015}).

\bibitem{GUL2015}
\bibinfo{author}{\mbox{\"O}. G{\"u}l} \emph{et~al.}
\newblock \bibinfo{title}{Towards high mobility {InSb} nanowire devices}.
\newblock \emph{\bibinfo{journal}{Nanotechnology}}
  \textbf{\bibinfo{volume}{26}}, \bibinfo{pages}{215202}
  (\bibinfo{year}{2015}).
  
\end{thebibliography}
\end{document}